\documentclass[twocolumn,superscriptaddress]{revtex4-2}
\usepackage{physics}
\usepackage{graphicx}
\usepackage{here}
\usepackage{amsmath}
\usepackage{amssymb}
\usepackage{bm}
\usepackage{wrapfig}
\usepackage{ascmac}
\usepackage{mathrsfs}
\usepackage{color}
\usepackage{comment}
\usepackage[normalem]{ulem}
\usepackage{amsthm}
\usepackage{mathtools}

\theoremstyle{plain}

\theoremstyle{definition}

\theoremstyle{remark}

\theoremstyle{plain}
\newtheorem*{thm*}{Theorem}
\newtheorem*{lem*}{Lemma}
\newtheorem*{prop*}{Proposition}
\newtheorem*{cond*}{Condition}
\newtheorem*{cor*}{Corollary}
\newtheorem*{conj*}{Conjecture}

\theoremstyle{definition}
\newtheorem*{ass*}{Assumption}
\newtheorem*{dfn*}{Definition}

\theoremstyle{remark}
\newtheorem*{rem*}{Remark}

\newcommand{\im}{{\rm i}}

\usepackage{comment}
\usepackage{hyperref}
\usepackage[normalem]{ulem}
\hypersetup{colorlinks=true,citecolor=blue,linkcolor=blue,urlcolor=blue}

\begin{document} 
\title{Entanglement asymmetry and quantum Mpemba effect in two-dimensional free-fermion systems} 
\author{Shion Yamashika}
\affiliation{Department of Physics, Chuo University, Bunkyo, Tokyo 112-8551, Japan} 
\affiliation{SISSA and INFN, via Bonomea 265, 34136 Trieste, Italy}
\author{Filiberto Ares}
\affiliation{SISSA and INFN, via Bonomea 265, 34136 Trieste, Italy}
\author{Pasquale Calabrese}
\affiliation{SISSA and INFN, via Bonomea 265, 34136 Trieste, Italy}
\affiliation{International Centre for Theoretical Physics (ICTP), Strada Costiera 11,
34151 Trieste, Italy}

\begin{abstract} 
The quantum Mpemba effect is the counter-intuitive non-equilibrium phenomenon wherein 
the dynamic restoration of a broken symmetry occurs more rapidly when the initial state exhibits a higher degree of symmetry breaking.
The effect has been recently discovered theoretically and observed experimentally in the framework of global quantum quenches, but so far it has only been investigated in one-dimensional systems.
Here we focus on a two-dimensional free-fermion lattice employing the entanglement asymmetry as a measure of symmetry breaking. 
Our investigation begins with the ground state analysis of a system featuring nearest-neighbor hoppings and superconducting pairings,
the latter breaking explicitly the $U(1)$ particle number symmetry. 
We compute analytically the entanglement asymmetry of a periodic strip using dimensional reduction, 
an approach that allows us to adjust the extent of the transverse size, achieving a smooth crossover between one and two dimensions.
Further applying the same method, we study the time evolution of the entanglement asymmetry after a quench to a Hamiltonian with only nearest-neighbor hoppings, preserving the particle number symmetry which is restored in the stationary state. 
We find that the quantum Mpemba effect is strongly affected by the size of the system in the transverse dimension, 
with the potential to either enhance or spoil the phenomenon depending on the initial states. 
We establish the conditions for its occurrence based on the properties of the initial configurations, extending the criteria found in the one-dimensional case.

\end{abstract}

\maketitle

\section{Introduction}
\label{sec:introduction}

Non-equilibrium physics usually defies our intuition, as we are 
generally used to near-equilibrium processes. One of the most 
counter-intuitive and puzzling out-of-equilibrium phenomena is the 
Mpemba effect: the farther a system is from equilibrium, the faster reaches it. Suppose for example that we wish to lower the temperature of a system immersed in a 
heat bath. If we gradually decrease the temperature 
of the bath, then the system passes by different 
near-equilibrium states until it reaches the desired temperature. 
Therefore, the larger the difference between the initial and final 
temperatures, the longer the time to cool the system. But what 
happens if we instantaneously change the temperature of the bath to 
the final value? Surprisingly, in this case, it may occur that 
the higher the initial temperature of the system is, the faster it 
cools down. 
Although the Mpemba effect was adverted even thousands of years ago, it 
has only recently been taken seriously~\cite{Mpemba-1969}. 
Nowadays we know that it occurs in a wide variety of systems and circumstances~\cite{Ahn-2016,Hu-2018, Chaddah-2010, Greaney-2011, Lasanta-2017, Keller-2018}. 
Its study has been specially boosted by the theoretical framework within Markovian 
dynamics developed in Ref.~\cite{Lu-2017} (see also~\cite{Klich-2019, Walker-2022, Teza-2023, Walker-2023, Bera-2023}), which has been proved experimentally 
in a controlled setting~\cite{Kumar-2020}, as well as for the reverse version~\cite{Kumar-2022}. However, numerous questions regarding when and why it occurs are still unanswered and subject to ongoing debate~\cite{Burridge-2016, Bechhoefer-2021}.   

It is quite natural to be curious about the occurrence of the Mpemba physics in quantum systems. 
Here two different research lines have surged. 
On the one hand, it has been analyzed in few-body open quantum systems after 
a quench of the temperature, both theoretically~\cite{Nava-2019,Kochsiek-2022,Carollo-2021,Manikandan-2021,Ivander-2023, Chatterjee-2023, Chatterjee-2023-2, Strachan-2024} and experimentally~\cite{Shapira-2024, Zhang-2024-Exp}. On the other hand, a 
phenomenon similar to the Mpemba effect has recently been found in isolated 
many-body quantum systems at zero temperature that evolve unitarily~\cite{Ares-2023}. More specifically, the system is prepared in a state 
that breaks a $U(1)$ symmetry and it is suddenly quenched to a 
Hamiltonian that respects the symmetry. The symmetry may not only be 
dynamically restored at large times in a subsystem 
but, unexpectedly, it may be faster restored when it is initially more broken. 
For generic one-dimensional free and interacting integrable systems, the origin and conditions under which this effect occurs have been understood in terms of the transport properties~\cite{Rylands-2023} exploiting the space-time duality approach~\cite{Bertini-2022, Bertini-2023, Bertini-2023-2} and employing the rule 54 cellular automaton and the Lieb-Liniger model as illustrative examples. 
Essentially, the quantum Mpemba effect occurs when, for the initial state that breaks more the symmetry, the charge is mainly carried by the fastest modes.
The mechanism behind the quantum Mpemba effect has been also studied in the XY spin chain~\cite{Murciano-2023}, where it is related to the Cooper pair content of the initial states.
Additionally, investigations have been conducted in the presence of gain and loss dissipation~\cite{Caceffo-2024}. 
Last, the quantum Mpemba effect has been experimentally observed in an ion trap that simulates a non-integrable model~\cite{Lata-2024}.
However, a theoretical understanding of the physical mechanism of the Mpemba effect for ergodic systems is still lacking.  

The basic tool for diagnosing the quantum Mpemba effect is the 
entanglement asymmetry, an observable based on the entanglement entropy that measures how much a symmetry is broken in a portion of 
an extended quantum system~\cite{Ares-2023}. Apart from characterizing the quantum Mpemba effect, the entanglement asymmetry provides new lenses to understand the non-equilibrium dynamics of 
extended quantum systems; for example, when the subsystem relaxes to
a non-Abelian Generalized Gibbs ensemble~\cite{Ares-2023-lack} and the symmetry is not restored, or when the symmetry is discrete~\cite{Ferro-2023,Capizzi-2023}. It has been also used to explore confinement phenomena in spin chains~\cite{Khor-2023}.
The properties of the entanglement asymmetry at equilibrium have been studied for generic compact Lie groups in matrix product states~\cite{Capizzi-Vitale-2023} and in critical systems described by a conformal 
field theory~\cite{Fossati-2024}, see also Ref.~\cite{Chen-2023}. In particular, in Ref.~\cite{Fossati-2024}, the entanglement asymmetry is connected with non-topological defects. 
The entanglement asymmetry of Haar random states breaking a $U(1)$ symmetry has been studied in the context of the information paradox of black hole evaporation~\cite{Ares-2023-bh}. 

\begin{figure}
    \centering
    \includegraphics[width=0.48\textwidth]{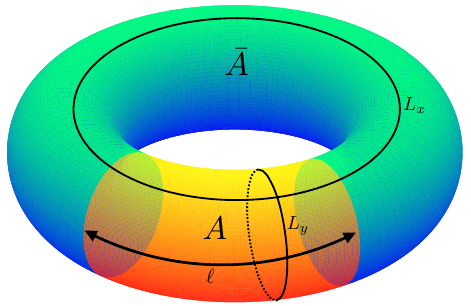}
    \caption{Schematic illustration of the 2D free fermionic system of size $L_x\times L_y$. 
    The terms longitudinal and transverse refer to the directions along $x$ and $y$, respectively. 
    Employing the entanglement asymmetry, we investigate the extent of particle number symmetry breaking in the periodic strip $A$ of width $\ell$. This analysis is conducted for both the ground state of~\eqref{eq:H_0} and after a sudden global quench to the Hamiltonian~\eqref{eq:H}.}
    \label{fig:system}
\end{figure}

So far the investigation of the quantum Mpemba effect has been restricted to one-dimensional (1D) systems. In the present manuscript, 
we wonder about the fate of this phenomenon when we add another spatial dimension. To facilitate the comparison with the known results, we take a setup analogous to the one studied in Ref.~\cite{Murciano-2023} in the 1D case: a two-dimensional (2D) lattice of 
free fermions with translationally invariant nearest-neighbor hoppings and superconducting terms. This system breaks the $U(1)$ particle number symmetry. 
We consider a 2D periodic lattice in both directions, i.e. a torus, and we 
take as a subsystem a periodic strip, as depicted in Fig.~\ref{fig:system}. 
This choice of the subsystem has the advantage that, since the configuration remains translationally invariant in one direction, the reduced density matrix factorizes into decoupled 1D factors. 
We can then exploit the exact knowledge of the entanglement asymmetry in the 1D configurations~\cite{Rylands-2023, Murciano-2023} to infer the 2D results. 
The dimensional reduction approach, introduced in Ref.~\cite{Chung-2000}, has been employed to study the (symmetry-resolved) entanglement entropy at equilibrium~\cite{Ares-2014, Murciano-2020} and, more recently, after quantum quenches~\cite{Yamashika-2023} (see also~\cite{Gibbins-2023}). 

Using dimensional reduction, we are going to derive exact analytic expressions first for the entanglement asymmetry in the ground state of the system with hoppings and superconducting pairings and then for its time evolution after a quench to a Hamiltonian with no pairing terms, which respects the particle number symmetry. 
As in 1D, the appearance of the quantum Mpemba effect hinges on the initial state that we choose, but here there is another crucial ingredient:
the size of the system in the transverse direction. 
By varying it, the quantum Mpemba effect can be either enhanced or spoiled depending on the initial configurations. 
Using the analytic results obtained for the entanglement asymmetry, we relate the occurrence of the Mpemba effect to the density of occupied modes in the initial state, extending the criteria found in 1D~\cite{Rylands-2023, Murciano-2023}.

The paper is organized as follows.  In Sec.\,\ref{sec:setup and definitions}, we introduce the setup and some basic quantities, including the entanglement asymmetry. 
In Sec.\,\ref{sec:dimensional reduction}, we describe how to calculate it in terms of the charged moments of the reduced density matrix using dimensional reduction. 
With this machinery, in Sec.\,\ref{sec:REA of gs}, we analyze the entanglement asymmetry at equilibrium and, in Sec.\,\ref{sec:time evolution of REA}, we examine its time evolution after the quench. In particular, we consider concrete initial states and show that the quantum Mpemba effect may occur. 
In Sec.\,\ref{sec:condition for QME}, we determine the microscopic conditions for its occurrence and give their physical interpretation. 
We finally summarize our results in Sec.\,\ref{sec:conclusion}.
We also include three appendices, where we derive some of the formulas presented in the main text.

\section{Setup and definitions}
\label{sec:setup and definitions}

We consider a system of free fermions on a 2D square lattice of size $L_x\times L_y$ described by the quadratic Hamiltonian 
\begin{multline}
    H_0
    = 
    -\frac{1}{2}
    \sum_{\mathbf{i}}\sum_{\nu=x,y}
    \qty(
    t_\nu 
    a_{\mathbf{i}+\mathbf{e}_\nu}^\dag 
    a_\mathbf{i}
    +\gamma_\nu 
    a_{\mathbf{i}+\mathbf{e}_\nu}^\dag 
    a_\mathbf{i}^\dag 
    +\mathrm{h.c.}
    )
    \\
    +
    h 
    \sum_\mathbf{i} 
    a_\mathbf{i}^\dag 
    a_\mathbf{i}, 
    \label{eq:H_0}
\end{multline}
where $\mathbf{i}=(i_x, i_y)$ labels a site of the lattice, 
$\mathbf{e}_x=(1,0)$ ($\mathbf{e}_y=(0,1)$) is the unit vector in the longitudinal (transverse) direction, and
$a_\mathbf{i}$ $(a_\mathbf{i}^\dag)$ is the annihilation (creation)
operator of a fermion at the site $\mathbf{i}$. The parameter $h$ is the chemical potential and $t_{x(y)}$, $\gamma_{x(y)}$ are the amplitudes of hopping and superconducting pairing in the longitudinal (transverse) direction, respectively. We impose periodic boundary conditions along both directions.  

For $\gamma_\nu\neq0$, the Hamiltonian~\eqref{eq:H_0} explicitly breaks the $U(1)$ symmetry 
associated with the particle number $Q=\sum_\mathbf{i}(a_\mathbf{i}^\dag a_\mathbf{i}-1/2)$, because $[H_0, Q]\neq 0$. 
Our main goal is to investigate the fate of this symmetry after a sudden global quench from the ground state $\ket{\mathrm{GS}}$ of the Hamiltonian~\eqref{eq:H_0} to the Hamiltonian
\begin{align}
    H = -\frac{1}{2}\sum_{\mathbf{i}}\sum_{\nu=x,y}
a_{\mathbf{i}+\mathbf{e}_\nu}^\dag a_\mathbf{i}
    +\mathrm{h.c.},
    \label{eq:H}
\end{align}
which commutes with $Q$ and, therefore, respects the $U(1)$ particle number symmetry. The same problem and setup have been considered in Ref.~\cite{Murciano-2023} in the 1D case, $L_y=1$. Here we will extend the results to $L_y>1$.

After the quench, the whole system evolves unitarily as $\ket{\Psi(t)}=e^{-\im t H}\ket{\rm GS}$. Therefore, it never globally relaxes and the $U(1)$ particle number symmetry is not restored at infinite time. However, we can focus on a subsystem $A$ like the periodic 
strip of width $\ell$ depicted in Fig.~\ref{fig:system}. 
This subsystem is described by the reduced density matrix
\begin{align}
    \rho_A(t)
    =
    \Tr_{\bar A}[
    \ket{\Psi(t)}\!\!\bra{\Psi(t)}],
\end{align}
where $\Tr_{\bar A}$ stands for the partial trace over $\bar A$, the complement of $A$. 
In the thermodynamic limit in the longitudinal direction $L_x\to\infty$, $\rho_A(t)$ is expected to relax into a stationary state at large times~\cite{Yamashika-2023, Rigol-2007, Barthel-2008, Cramer-2008, Calabrese-2012, Fagotti-2013, Polkovnikov-2011,Calabrese-2016,Vidmar-2016,Essler-2016,Doyon-2020,Bastianello-2022,Alba-2021} that generically respects the $U(1)$ particle number symmetry. In other words, if $Q_A=\sum_{\mathbf{i}\in A}(a_\mathbf{i}^\dag a_\mathbf{i}-1/2)$ is the restriction of $Q$ to the subsystem $A$, the reduced density matrix $\rho_A(t)$ is expected to satisfy $\lim_{t\to\infty}[\rho_A(t), Q_A]=0$. 

To monitor the time evolution of the $U(1)$  symmetry in $A$, we can employ the R\'enyi entanglement asymmetry~\cite{Ares-2023}, which quantifies how much the symmetry is broken in the subsystem. This observable is defined as 
\begin{align}
    \Delta S_A^{(n)}
    = 
    S_n[\rho_{A,Q}(t)]-S_n[\rho_A(t)], 
    \label{eq:def_of_REA}
\end{align}
where $S_n[\rho]=\log(\Tr[\rho^n])/(1-n)$ is the R\'enyi entropy of $\rho$. The matrix $\rho_{A,Q}(t)$ is the symmetrization of $\rho_A(t)$ and is defined as $\rho_{A,Q}(t)=\sum_{q} \Pi_q \rho_A(t) \Pi_q$, with $\Pi_q$ the projection operator onto the eigenspace of $Q_A$ with eigenvalue $q$. 
Taking the limit $n\to1$ in Eq.\,\eqref{eq:def_of_REA},  we obtain the von Neumann entanglement asymmetry, which is the relative entropy between $\rho_{A}(t)$ and $\rho_{A,Q}(t)$,
\begin{align}
    \Delta S_A^{(1)}
    := 
    \Tr[
    \rho_A(t)
    (
    \log \rho_A(t)
    -\log \rho_{A,Q}(t)
    )
    ]. 
\end{align}
The R\'enyi entanglement asymmetry is definite positive, $\Delta S_A^{(n)}\geq0$, and vanishes if and only if $[\rho_A(t),Q_A]=0$, i.e., $\rho_A(t)$ is symmetric~\cite{Ma-2022, Han-2023}. For integer $n\geq 2$, it can be obtained experimentally using the randomized
measurement toolbox as  done in~\cite{Lata-2024} in an ion trap.

\subsection{Diagonalization of the Hamiltonian}
Here we diagonalize the Hamiltonian~\eqref{eq:H_0} and determine its ground state as the first step in the analysis of the entanglement asymmetry. Readers who are familiar with the diagonalization of quadratic fermionic Hamiltonians using Bogoliubov rotations can skip this section.
\par 
Since the Hamiltonian~\eqref{eq:H_0} is quadratic and translationally invariant in both 
directions, we introduce the Fourier modes
\begin{align}
    a_\mathbf{q}
    =
    \frac{1}{\sqrt{L_xL_y}}
    \sum_\mathbf{i}
    e^{-\im \mathbf{q\cdot i}}
    a_\mathbf{i},
    \label{eq:Fourier}
\end{align}
with quasi-momenta $q_x=2\pi n_x/L_x$ ($n_x=0,1,..,L_x-1$) and $q_y=2\pi n_y/L_y$ ($n_y=0,1,..,L_y-1$). In the Fourier space, the 
Hamiltonian~\eqref{eq:H_0} reads
\begin{align}
    H_0
    = 
    \frac{1}{2}
    \sum_\mathbf{q}
    \qty(
    \xi_\mathbf{q}
    a_\mathbf{q}^\dag 
    a_\mathbf{q}
    +
    \im 
    \Delta_\mathbf{q}
    a_\mathbf{q}^\dag 
    a_\mathbf{-q}^\dag
    +\mathrm{h.c.}
    ),
    \label{eq:H_0_Fourier}
\end{align}
where $\xi_\mathbf{q}=h-t_x \cos q_x-t_y\cos q_y$ and $\Delta_\mathbf{q}=\gamma_x\sin q_x+\gamma_y\sin q_y$. 
As well-known, the Hamiltonian~\eqref{eq:H_0_Fourier} can be diagonalized by the Bogoliubov transformation \cite{Lieb-1961}
\begin{align}
    \mqty(
    \eta_\mathbf{q}
    \\
    \eta_{-\mathbf{q}}^\dag
    )   
    = 
    \mqty(
    \cos \frac{\phi_\mathbf{q}}{2}
    &
    \im 
    \sin \frac{\phi_\mathbf{q}}{2}
    \\
    \im 
    \sin \frac{\phi_\mathbf{q}}{2}
    & 
    \cos \frac{\phi_\mathbf{q}}{2}
    )
    \mqty(
    a_\mathbf{q}
    \\
    a_{-\mathbf{q}}^\dag
    ), 
    \label{eq:Bogoliubov_transform}
\end{align}
where $\phi_\mathbf{q}=\arctan(\Delta_\mathbf{q}/\xi_\mathbf{q})+\pi\Theta(-\xi_\mathbf{q})$, with $\Theta(x)$ being the Heviside step function.
Using Eq.~\eqref{eq:Bogoliubov_transform} in Eq.~\eqref{eq:H_0_Fourier}, we obtain 
\begin{align}
    H_0 
    = 
    \sum_\mathbf{q}
    \sqrt{
    \xi_\mathbf{q}^2
    +
    \Delta_\mathbf{q}^2 
    }
    \left(\eta_\mathbf{q}^\dag 
    \eta_\mathbf{q}-\frac{1}{2}\right).
\end{align}
Therefore, the ground state $\ket{\mathrm{GS}}$ of the Hamiltonian \eqref{eq:H_0} is the vacuum state annihilated by $\eta_\mathbf{q}$ for all $\mathbf{q}$. 
From this condition, $\eta_\mathbf{q}\ket{\mathrm{GS}}=0$, we obtain its explicit form,
\begin{align}
    \ket{\mathrm{GS}}
    = 
    \bigotimes_{n_x=0}^{L_x-1}
    \ket{\mathrm{GS}_{q_x}}, 
    \label{eq:Psi_0}
\end{align}
where 
\begin{align}
    \ket{\mathrm{GS}_{q_x}}
    =
    \bigotimes_{n_y=0}^{\frac{L_y}{2}-1}
    \qty(
    \cos \frac{\phi_\mathbf{q}}{2}
    -\im 
    \sin \frac{\phi_\mathbf{q}}{2}
    a_\mathbf{q}^\dag 
    a_\mathbf{-q}^\dag
    )\ket{0_\mathbf{q}},
    \label{eq:Psi_0_q}
\end{align}
and $\ket{0_\mathbf{q}}$ is the fermionic vacuum state that is annihilated by $a_\mathbf{\pm q}$, i.e., $a_\mathbf{\pm q}\ket{0_\mathbf{q}}=0$. 
Eq.~\eqref{eq:Psi_0_q} implies that Cooper pairs consisting of fermions with opposite momenta are condensed in $\ket{\mathrm{GS}}$ in analogy to the BCS wavefunction of superconductors~\cite{Bardeen-1957}. 
The emergence of Cooper pairs is induced by the superconducting pairing terms in Eq.\,\eqref{eq:H_0} and implies that $\ket{\mathrm{GS}}$ breaks the $U(1)$ particle number symmetry.  

In terms of the Fourier modes~\eqref{eq:Fourier}, the post-quench 
Hamiltonian~\eqref{eq:H} is diagonal
\begin{align}
    H = \sum_\mathbf{q} \epsilon_\mathbf{q}a_\mathbf{q}^\dag a_\mathbf{q},
    \label{Hdiag2}
\end{align}
where $\epsilon_\mathbf{q}=-\cos q_x-\cos q_y$ is the single-particle dispersion relation. 

\section{Charged moments and dimensional reduction} 
\label{sec:dimensional reduction}

In this section, we describe the methods to calculate
the R\'enyi entanglement asymmetry both in the ground state of the 
Hamiltonian~\eqref{eq:H_0} and after the quench to~\eqref{eq:H}. 
Readers who are familiar with the calculation of the entanglement asymmetry in fermionic Gaussian states and dimensional reduction or are only interested in the main results can skip this section.

\par 
Let us first relate the R\'enyi entanglement asymmetry~\eqref{eq:def_of_REA} to the charged moments of $\rho_A(t)$. Using the Fourier representation of the projection operator $\Pi_q$, 
\begin{align}
    \Pi_q 
    = 
    \int_0^{2\pi}
    \frac{d\alpha}{2\pi}
    e^{\im \alpha(Q_A-q)},
    \label{eq:Pi_q Fourier}
\end{align}
the symmetrized reduced density matrix $\rho_{A,Q}(t)$ can be written as 
\begin{align}
    \rho_{A,Q} (t)
    = 
    \int_0^{2\pi}
    \frac{d\alpha}{2\pi}
    e^{\im \alpha Q_A} 
    \rho_A(t)
    e^{-\im \alpha Q_A}.
    \label{eq:rho_AQ_Fourier} 
\end{align}
Plugging it into Eq.\,\eqref{eq:def_of_REA}, we obtain 
\begin{align}
    \Delta S_A^{(n)}
    = 
    \frac{1}{1-n}
    \log
    \left[\int \limits_{[0,2\pi]^n}
    \frac{d^n \boldsymbol{\alpha}}{(2\pi)^n}
    \frac{Z_n(\boldsymbol{\alpha},t)}{Z_n(\boldsymbol{0},t)}\right],
    \label{eq:REA_Zn}
\end{align}
where $Z_n(\boldsymbol{\alpha},t)$ are the charged moments 
\begin{align}
    Z_n(\boldsymbol{\alpha},t)
    = 
    \Tr 
    \left[\prod_{j=1}^n 
    \rho_A(t) e^{\im \alpha_{j,j+1}Q_A}\right], 
    \label{eq:charged_moment}
\end{align}
and $\alpha_{j,j+1}=\alpha_j-\alpha_{j+1}$ with $\alpha_{n+1}=\alpha_1$. 

The state $\ket{\Psi(t)}$ after the quench
satisfies Wick theorem (because it is a free evolution of a Gaussian initial state) and, consequently, the reduced density matrix $\rho_A(t)$ is Gaussian. 
Hence the time-evolved reduced density matrix $\rho_A(t)$ is univocally determined by the 
restriction to $A$ of the two-point correlation matrix~\cite{Peschel-2003}
\begin{equation}\label{eq:two_poin_corr}
\Gamma_{\mathbf{i}, \mathbf{i}'}(t)=2\bra{\Psi(t)}\mathbf{a}_{\mathbf{i}}^\dagger \mathbf{a}_{\mathbf{i}'}\ket{\Psi(t)}-\delta_{\mathbf{i}, \mathbf{i}'},
\end{equation}
with $\mathbf{a}_{\mathbf{i}}=(a_{\mathbf{i}}^\dagger, a_{\mathbf{i}})$ and $\mathbf{i},\mathbf{i}'\in A$. This is a $2V_A\times 2V_A$ matrix, where $V_A=\ell L_y$ is the volume of subsystem $A$. 

Note that the operators $e^{i\alpha_{jj+1} Q_A}$ that enter in Eq.~\eqref{eq:charged_moment} are also Gaussian. 
Thus the charged moments $Z_n(\boldsymbol{\alpha}, t)$ are the trace of a product of Gaussian operators. 
Applying the composition rules of Gaussian operators~\cite{Balian-1969, Fagotti-2010}, as described in detail in Appendix \ref{app:charged moment}, we find that 
\begin{equation}\label{eq:2d_charged_moments_corr_mat}
Z_n(\boldsymbol{\alpha},t)
= 
\sqrt{\det\qty[\qty(\frac{I-\Gamma(t)}{2})^{n}(I+W_n(\boldsymbol{\alpha},t))]},
\end{equation}
where 
\begin{equation}{\label{eq:def of W_n}}
W_n(\boldsymbol{\alpha}, t)=\prod_{j=1}^n\frac{I+\Gamma(t)}{I-\Gamma(t)}e^{\im \alpha_{j, j+1}I\otimes\sigma_z}.
\end{equation}
Note that the expression \eqref{eq:2d_charged_moments_corr_mat} for the charged moments is exact and can be applied to any subsystem geometry. 
\par 
Given that the system is translationally invariant in both directions and the subsystem $A$ is periodic in the $y$ axis, the charged moments can be further factorized into the ones of $L_y$
decoupled 1D chains~\cite{Chung-2000,Yamashika-2023}. To this end,
it is convenient to introduce creation/annihilation fermionic operators in the mixed space-momentum basis by taking the partial Fourier transform in the transverse direction, 
\begin{align}
    c_{i_x,q_y}
    = 
    \frac{1}{\sqrt{L_y}}
    \sum_{i_y=0}^{L_y-1}
    e^{-\im q_yi_y}
    a_{i_x,i_y}.
\end{align}
In terms of them, the entries of the two-point correlation 
matrix~\eqref{eq:two_poin_corr} read ($q_y=2\pi n_y/L_y$)
\begin{equation}\label{eq:Gamma_block_diagonal}
\Gamma_{(i_x, j_y), (i_x', j_y')}=\frac{1}{L_y}\sum_{n_y=0}^{L_y-1}
e^{\im \frac{2\pi n_y}{L_y}(j_y-j_y')}(\Gamma_{n_y})_{i_x, i_x'},
\end{equation}
where $\Gamma_{n_y}$ are $2\ell\times 2\ell$ matrices that, in the thermodynamic limit $L_x\to\infty$, are block Toeplitz and take the form
\begin{equation}
    (\Gamma_{n_y}(t))_{i_x,i_x'}
    = 
    \int_0^{2\pi}
    \frac{dq_x}{2\pi}
    e^{-\im q_x(i_x-i_x')}
    \mathcal{G}_{(q_x, q_y)}(t),
    \label{eq:Gamma_ny}
\end{equation}
with $2\times 2$ symbol
\begin{equation}
\mathcal{G}_{(q_x, q_y)}(t)=
\mqty(
    \cos \phi_\mathbf{q} 
    & 
    -\im e^{-2\im t \epsilon_\mathbf{q}}\sin \phi_\mathbf{q}
    \\
    \im e^{2\im t \epsilon_\mathbf{q}}\sin \phi_\mathbf{q}
    & 
    -\cos \phi_\mathbf{q} 
    ). 
\end{equation}
Notice that the dependence on $n_y$ is encoded in the transverse momentum $q_y=2\pi n_y/L_y$.

Eq.~\eqref{eq:Gamma_block_diagonal} shows that the two-point 
correlation matrix $\Gamma(t)$ is block diagonal in the transverse 
momentum sectors, labelled by $n_y$, due to the translational 
symmetry of $\ket{\Psi(t)}$ in that direction~\cite{Chung-2000}. 
This implies that the reduced density matrix $\rho_A(t)$ factorizes 
as
\begin{align}
    \rho_A(t) 
    = 
    \bigotimes_{n_y=0}^{L_y-1}
    \rho_{A,n_y}(t), 
    \label{eq:rho_A_product}
\end{align}
where $\rho_{A,n_y}(t)$ is the reduced density matrix of a single interval of length $\ell$ in an infinite 1D free fermionic chain, 
univocally determined by the two-point correlation matrix 
$\Gamma_{n_y}(t)$, introduced in Eq.~\eqref{eq:Gamma_ny}.  
By simple algebra, we find that the operators $e^{\im \alpha Q_A}$ 
in Eq.\,\eqref{eq:charged_moment} can also be written as 
\begin{align}
    e^{\im \alpha Q_A} 
    = 
    \bigotimes_{n_y=0}^{L_y-1}
    e^{\im \alpha Q_{A,n_y}}, 
    \label{eq:expQ_product}
\end{align}
where $Q_{A,n_y}= \sum_{i_x\in A} \mathbf{c}_{i_x,q_y} \sigma_z \mathbf{c}_{i_x,q_y}^\dag /2$ with $\mathbf{c}_{i_x, q_y}=(c_{i_x, q_y}^\dag, c_{i_x, q_y})$. 
Plugging Eqs.~\eqref{eq:rho_A_product}~and~\eqref{eq:expQ_product} into Eq.~\eqref{eq:charged_moment}, 
the charged moments factorize into the transverse momentum sectors as
\begin{align}
    Z_n(\boldsymbol{\alpha},t)
    = 
    \prod_{n_y=0}^{L_y-1}
    Z_{n,n_y}(\boldsymbol{\alpha},t),
    \label{eq:Z_n_product}
\end{align}
where 
\begin{align}
    Z_{n,n_y}(\boldsymbol{\alpha},t)
    = 
    \Tr 
    \left[\prod_{j=1}^{n}\rho_{A,n_y}(t) e^{\im \alpha_{j,j+1} Q_{A,n_y}}\right] 
\end{align}
are the charged moments of the 1D state $\rho_{A,n_y}(t)$. 
\par 
Since $\rho_{A,n_y}(t)$ is determined by the two-point correlation matrix $\Gamma_{n_y}(t)$, its charged moments can be calculated with a formula similar to Eq.~\eqref{eq:2d_charged_moments_corr_mat}, replacing $\Gamma(t)$ by $\Gamma_{n_y}(t)$,
\begin{align}
    Z_{n,n_y}(\boldsymbol{\alpha},t)
    = 
    \sqrt{\det[
    \qty(\frac{I-\Gamma_{n_y}(t)}{2})^n 
    (I+W_{n,n_y}(\boldsymbol{\alpha},t))
    ]},
    \label{eq:Z_nny_det}
\end{align}
where
\begin{align}
    W_{n, n_y}(\boldsymbol{\alpha},t)=\prod_{j=1}^{n}
    \frac{I+\Gamma_{n_y}(t)}{I-\Gamma_{n_y}(t)}
    e^{\im \alpha_{j,j+1} I\otimes \sigma_z}.
\end{align}
As $Z_{n,n_y}(\boldsymbol{\alpha},t)$ are the charged moments of the 1D state $\rho_{A,n_y}(t)$, we can get exact analytic expressions for them applying the results for 1D free fermionic systems obtained in Refs.~\cite{Rylands-2023, Murciano-2023}. 
In what follows, we will combine them with the machinery developed in this section to compute first the entanglement asymmetry in the ground state of the Hamiltonian~\eqref{eq:H_0}, Sec.~\ref{sec:REA of gs}, and then its time evolution after a quench to the Hamiltonian~\eqref{eq:H}, Sec.~\ref{sec:time evolution of REA}. 
 
\section{Entanglement asymmetry in the ground state of the pairing Hamiltonian}
\label{sec:REA of gs}

In this section, we study the entanglement asymmetry of the periodic strip $A$ in the ground state~\eqref{eq:Psi_0} of the Hamiltonian~\eqref{eq:H_0}. This will be the initial state in the quench protocol. 

As we have seen in the previous section, the R\'enyi entanglement 
asymmetry can be calculated from the charged moments of $\rho_A$.
According to Eq.~\eqref{eq:Z_n_product}, they factorize in the charged moments of an interval $\ell$ of $L_y$ decoupled infinite 1D fermionic chains in the 
states described by the two-point correlation functions of 
Eq.~\eqref{eq:Gamma_ny}. The analytic expression of 
the charged moments for a large interval $\ell$ in this kind of 1D chains has been obtained in 
Ref.~\cite{Murciano-2023}. It reads 
\begin{equation}
\frac{Z_{n, n_y}(\boldsymbol{\alpha})}{Z_{n, n_y}(\boldsymbol{0})}=e^{\ell A_{n, n_y}(\boldsymbol{\alpha})},
\end{equation}
where
\begin{equation}
A_{n, n_y}(\boldsymbol{\alpha})=
\int_{0}^{2\pi}\frac{dq_x}{4\pi} \sum_{j=1}^n f_{\boldsymbol{q}}(\alpha_{j,j+1})
\end{equation}
and
\begin{equation}\label{eq:f_q}
f_{\mathbf{q}}(\alpha)=\log(\cos\alpha+{\rm i}\cos\phi_{\mathbf{q}}\sin\alpha).
\end{equation}
Therefore, using Eq.~\eqref{eq:Z_n_product},
the charged moments in the ground state of the 
Hamiltonian~\eqref{eq:H_0} for the periodic strip of Fig.~\ref{fig:system} behave when $L_x\to\infty$ and $\ell$ is large as 
\begin{equation}
\frac{Z_n(\boldsymbol{\alpha})}{Z_n(\boldsymbol{0})}=e^{\ell\sum_{n_y=0}^{L_y-1}A_{n, n_y}(\boldsymbol{\alpha})}.
\label{eq:charged_moment_t=0}
\end{equation}
Inserting this result in Eq.~\eqref{eq:REA_Zn}, we can derive the 
R\'enyi entanglement asymmetry. For $\ell\gg 1$, the 
$n$-fold integral in Eq.~\eqref{eq:REA_Zn} can be solved by 
performing a saddle point approximation as we describe in Appendix~\ref{app:saddle_point}. We eventually obtain
\begin{align}
    \Delta S_A^{(n)}
    \simeq 
    \frac{1}{2}
    \log
    \frac{V_A\varrho_\mathrm{c} n^{\frac{1}{n-1}}\pi}{4},
    \label{eq:REA_t=0}
\end{align}
where 
\begin{equation}
\varrho_\mathrm{c}=\frac{1}{L_y}\sum_{n_y=0}^{L_y-1}\int_0^{2\pi}\frac{d q_x}{2\pi}\sin^2\phi_{\mathbf{q}}.
\label{eq:Cooper_pair_density}
\end{equation}
In the von Neumann limit $n\to1$, Eq.~\eqref{eq:REA_t=0} reduces to
\begin{align}\label{eq:EA_t=0_vN_limit}
    \Delta S_A^{(1)}
    \simeq 
    \frac{1}{2}
    \log 
    \frac{V_A \varrho_{\rm c}\pi}{4}+\frac{1}{2}.
\end{align}
Since $\sin^2\phi_{\mathbf{q}}=|\bra{{\rm GS}}a_{\mathbf{q}}a_{-\mathbf{q}}\ket{{\rm GS}}|^2$ is the mode occupation of Cooper pairs,
$\varrho_\mathrm{c}$ is the density of Cooper 
pairs in the ground state. Therefore, Eq.~\eqref{eq:REA_t=0} shows 
that the R\'enyi entanglement asymmetry increases logarithmically 
with the density of Cooper pairs $\varrho_\mathrm{c}$ that the state 
contains. This result is quite natural because, as we have already 
mentioned, the condensation of Cooper pairs originates from the 
explicit breaking of the $U(1)$ particle number symmetry in the 
Hamiltonian~\eqref{eq:H_0} \cite{Bardeen-1957}. Equation \eqref{eq:REA_t=0}
generalizes to $L_y>1$ the result 
for the ground state of $\eqref{eq:H_0}$ in the 1D case $L_y=1$~\cite{Murciano-2023}. 
Equation \eqref{eq:REA_t=0} is compatible with the results of Ref.~\cite{Capizzi-Vitale-2023}, which indicate that the entanglement asymmetry for generic compact Lie groups 
grows with the logarithm of the subsystem 
size. The prefactor is proportional to 
half of the Lie group dimension, which for $U(1)$ is precisely $1/2$, 
see also Ref.~\cite{Fossati-2024}.

\begin{figure}
    \includegraphics[width=0.5\textwidth]{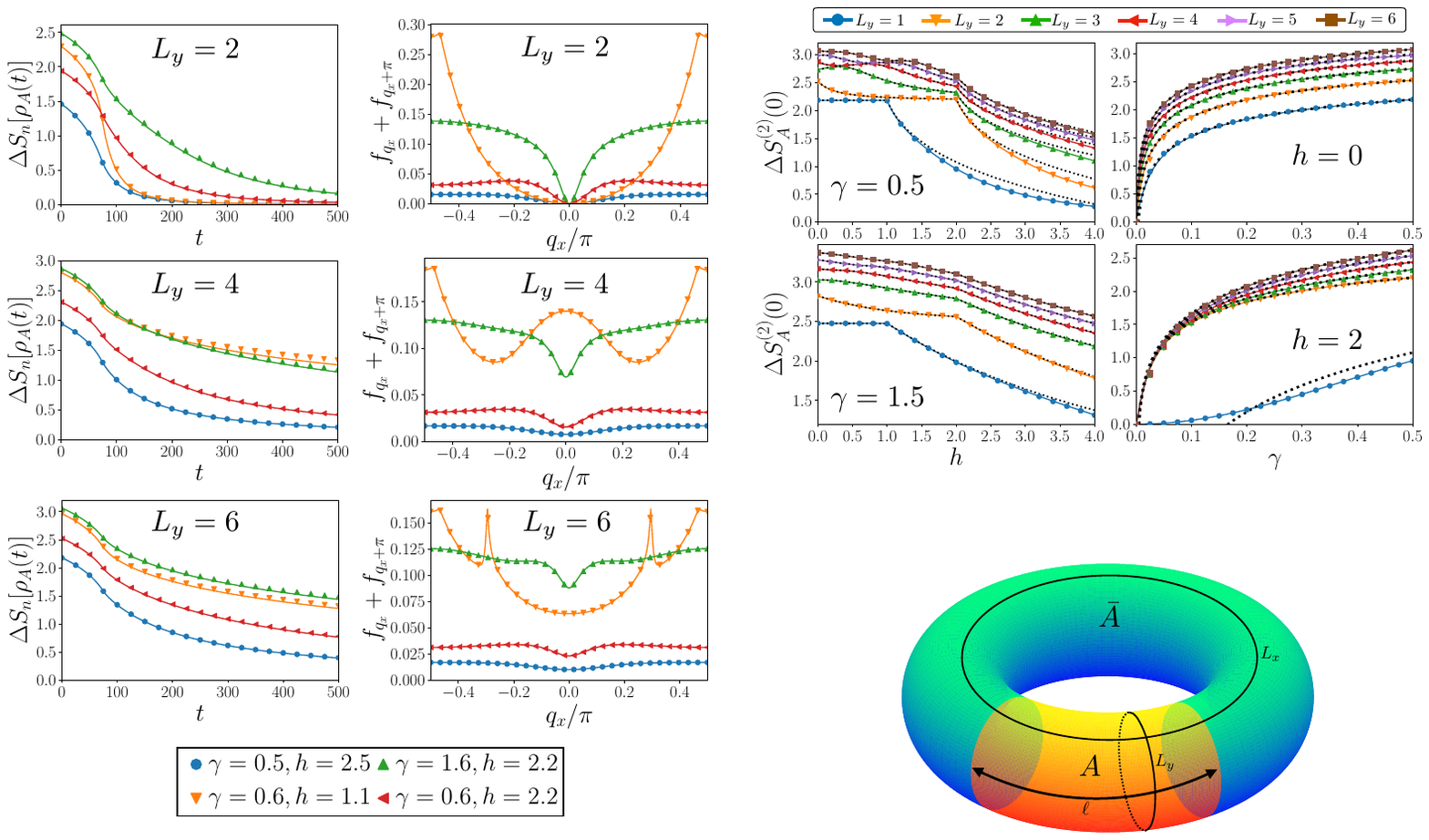}
    \caption{R\'enyi entanglement asymmetry for R\'enyi index $n=2$ in the ground state of the Hamiltonian~\eqref{eq:H_0} with $t_x=t_y=1$ and $\gamma_x=\gamma_y=\gamma$. We take as subsystem $A$ the periodic strip of Fig.~\ref{fig:system}, with $\ell=100$ and different lengths $L_y$ of the transverse dimension. The symbols are the exact value computed numerically using Eq.~\eqref{eq:Z_nny_det}, the solid lines have been obtained from the analytic expression of the charged moments in Eq.~\eqref{eq:charged_moment_t=0} and the dotted lines are the saddle point approximation of Eq.~\eqref{eq:REA_t=0}. 
    }
    \label{fig:REA_isotropic}
\end{figure}

In Fig.~\ref{fig:REA_isotropic}, we check the validity of the saddle point approximation~\eqref{eq:REA_t=0} for the case of isotropic pairing, $\gamma_x=\gamma_y=\gamma$, and hopping, $t_x=t_y=1$, by comparing it with the exact result obtained by evaluating numerically the charged moments using Eq.~\eqref{eq:Z_nny_det}.  We plot the R\'enyi entanglement asymmetry for $n=2$ as a function of $h$ and fixed $\gamma$ (left panels) and vice-versa (right panels) taking a strip of width $\ell=100$ and different values for the transverse dimension $L_y$. 
We observe in general a good agreement between Eq.~\eqref{eq:REA_t=0} (dashed lines) and the exact result (symbols), except for $h>1$ and $L_y$  small; in the latter case  $\ell$ must be large enough to find a good agreement. 
We also plot the R\'enyi entanglement asymmetry obtained by integrating numerically in $\boldsymbol{\alpha}$ the analytic expression for the charged moments in Eq.\,\eqref{eq:charged_moment_t=0}, instead of using the saddle point approximation.
The resulting curves (solid lines) perfectly match the exact result in all the cases considered. 
We observe that the R\'enyi entanglement asymmetry decreases as $h$ increases and when $\gamma$ decreases. 
The explanation lies in the fact that the Hamiltonian \eqref{eq:H_0} becomes symmetric for $|h|\to \infty$ and when $\gamma_x=\gamma_y=0$. 
\par 
A peculiar feature emerging neatly from Fig.~\ref{fig:REA_isotropic} at fixed $\gamma$ is the presence of a non-analyticity in $h$.
For $L_y=1$ and fixed pairing $\gamma$ (left panels), the asymmetry is constant for $h<1$ and varies for $h>1$. 
In fact, $h=1$ is the critical point where the mass gap vanishes. 
For all $L_y>1$, the non-analytic point moves to $h=2$ and it is slowly smoothed as $L_y$ increases. Indeed, the critical point is now at $h=2$ for all $L_y>1$, as clear from the dispersion relation in Eq.~\eqref{Hdiag2}. 
We can understand why the non-analiticity smears as $L_y$ increases from Eqs.~\eqref{eq:REA_t=0} and~\eqref{eq:Cooper_pair_density}. According to them, the entanglement asymmetry is given by the Cooper pair density $\varrho_{\rm c}$, which decomposes into the sum of the contributions of $L_y$ 1D fermionic chains labeled by the transverse momentum sectors $n_y=0, 1, \dots, L_y-1$ and with the single-particle dispersion relation
$\omega_{q_x, n_y}=\sqrt{\xi_{q_x, 2\pi n_y/L_y}^2+\Delta_{q_x, 2\pi n_y/L_y}^2}$. Observe that, at the critical point $h=2$, all these 1D chains have a non-zero mass gap except for the one corresponding to $n_y=0$. The non-analyticity of the entanglement asymmetry at $h=2$ comes
from the contribution of the $n_y=0$ gapless 1D chain, while the rest of them give a smooth contribution since they are gapped. For small $L_y$, the term $n_y=0$ in Eq.~\eqref{eq:REA_t=0} is dominant and its non-analiticity at $h=2$ is well visible. As we increase $L_y$, the smooth contribution from the terms $n_y\neq 0$ becomes dominant, hiding the non-analyticity.
%

\section{Time evolution of the entanglement asymmetry after the quench}
\label{sec:time evolution of REA}

In this section, we apply the dimensional reduction approach of Sec.~\ref{sec:dimensional reduction} to determine the time evolution of the R\'enyi entanglement asymmetry in the periodic strip $A$ after a quench to the Hamiltonian~\eqref{eq:H} from the ground state of the Hamiltonian~\eqref{eq:H_0}. We then examine in detail the quenches from some specific physical choices of the parameters of the initial Hamiltonian~\eqref{eq:H_0}.

\subsection{Quasi-particle picture for the charged moments and the entanglement asymmetry}

In Refs.~\cite{Rylands-2023, Murciano-2023}, the time evolution of the charged moments has been studied for an interval $\ell$ of an infinite 1D free-fermionic chain in a state as the ones in which our reduced density matrix factorizes in Eq.~\eqref{eq:Z_n_product}. 
The direct application of the results obtained in those works tells us that the charged moments associated with the $n_y$ longitudinal mode behave, in the ballistic regime $\ell, t\to\infty$, with $\zeta=t/\ell$ fixed and $L_x\to\infty$, as
\begin{equation}\label{eq:time_ev_ny_charged_mom}
\frac{Z_{n, n_y}(\boldsymbol{\alpha}, t)}{Z_{n, n_y}(\boldsymbol{0}, t)}=e^{\ell A_{n, n_y}(\boldsymbol{\alpha}, \zeta)},
\end{equation}
where
\begin{equation}\label{eq:A_n_ny_t}
A_{n, n_y}(\boldsymbol{\alpha}, \zeta)=\int_{0}^{2\pi}\frac{dq_x}{4\pi}x_{q_x}(\zeta)\sum_{j=1}^n f_{\mathbf{q}}(\alpha_{j, j+1}).
\end{equation}
In this expression, $x_{q_x}(\zeta)=1-\min(2\zeta|v_{q_x}|, 1)$ and $v_{q_x}$ is the group velocity of the excitations of the post-quench Hamiltonian in the longitudinal direction, $v_{q_x}=\partial_{q_x}\epsilon_{\mathbf{q}}$. The function $f_{\mathbf{q}}(\alpha)$ is defined in Eq.~\eqref{eq:f_q}. 

Combining Eqs.~\eqref{eq:time_ev_ny_charged_mom} and \eqref{eq:Z_n_product}, the charged moments for the periodic strip
$A$ evolve after the quench as
\begin{equation}\label{eq:2D_time_ev_charged_moments}
\frac{Z_n(\boldsymbol{\alpha}, t)}{Z_n(\boldsymbol{0}, t)}=e^{\ell \sum_{n_y=0}^{L_y-1}A_{n, n_y}(\boldsymbol{\alpha}, \zeta)}.
\end{equation}
This expression can be interpreted in the light of the quasi-particle 
picture, originally developed for the entanglement entropy in 1D integrable models~\cite{Calabrese-2005, Alba-2017,ac-17b,c-18} and recently also extended to 2D free fermionic systems~\cite{Yamashika-2023, Gibbins-2023}. 
Since $Z_n(\boldsymbol{\alpha}, t)$ factorizes into the charged moments of a single interval of decoupled 1D free fermionic chains, we
can readily apply the quasi-particle interpretation elaborated in Refs.~\cite{Ares-2023, Murciano-2023, Caceffo-2024} for the charged moments and the entanglement asymmetry in 1D. In each decoupled 1D chain, the quench generates uniformly pairs of entangled excitations
that propagate ballistically with opposite momentum $\pm q_x$ and velocity $v_{q_x}$. Since the correlator $\langle a_{\mathbf{q}}a_{-\mathbf{q}}\rangle$ is not zero, they form a Cooper pair that contributes to the breaking of the $U(1)$ symmetry in the interval $\ell$ and, therefore, to the entanglement asymmetry as long as both excitations are inside the interval. Applying this interpretation in Eq.~\eqref{eq:A_n_ny_t}, the function $\ell x_{q_x}(\zeta)$ counts the number of Cooper pairs with momentum $|q_x|$ inside an interval of length $\ell$ and $\sum_jf_{\mathbf{q}}(\alpha_{j, j+1})$ quantifies the contribution of each pair to the charged moments. We refer the reader to Ref.~\cite{Caceffo-2024} for a full-fledge description of 1D charged moments and entanglement asymmetry using the quasi-particle picture that also encompasses the case of multiplets. 

Inserting~\eqref{eq:2D_time_ev_charged_moments} in Eq.~\eqref{eq:REA_Zn}, we can determine
the time evolution of the R\'enyi entanglement asymmetry.
According to Eq.~\eqref{eq:A_n_ny_t}, 
the time-evolved 
charged moments $Z_n(\boldsymbol{\alpha}, t)$ not only factorize in the transverse momentum modes $n_y$
but also in the number of replicas $n$. Exploiting this property, the 
R\'enyi entanglement asymmetry can be written in the form
\begin{equation}\label{eq:time_ev_REA_Jk}
\Delta S_A^{(n)}(t)=\frac{1}{1-n}\log\left[\sum_{k=-\infty}^\infty J_k(t)^n\right],
\end{equation}
with
\begin{equation}\label{eq:J_k}
J_k(t)=\int_{0}^{2\pi}\frac{d\alpha}{2\pi}e^{{\rm i}\alpha k} e^{\ell \sum_{n_y=0}^{L_y-1} A_{1, n_y}(\alpha, \zeta)},
\end{equation}
as we show in Appendix~\ref{app:vN limit}, generalizing the results in 1D~\cite{Rylands-2023}. In 
Eq.~\eqref{eq:time_ev_REA_Jk}, we can 
analytically continue the R\'enyi index $n$ to real
values and then take the replica limit $n\to 1$. We find, see Appendix~\ref{app:vN limit},
\begin{equation}\label{eq:time_ev_EA_Jk}
\Delta S_A^{(1)}(t)=-\sum_{k=-\infty}^{\infty}
\Re[J_k(t)\log J_{k}(t)].
\end{equation}

We will now consider the evolution of the entanglement asymmetry after the quench to~\eqref{eq:H} starting from the ground state of the Hamiltonian~\eqref{eq:H_0} for 
different values of the parameters $t_x$, $t_y$, $\gamma_x$, $\gamma_y$ and $h$.
We will use the analytic expressions that we have just obtained, which we will check
against the exact values computed numerically using Eq.~\eqref{eq:Z_nny_det}. 

\subsection{Partially-filled product states}
\label{subsec:PF}

In the space of parameters of the Hamiltonian~\eqref{eq:H_0}, a relevant subset for analyzing the $U(1)$ particle number symmetry is constituted by those whose ground state is a \textit{partially-filled} product state built from the 1D cat state 
\begin{align}
    \ket{\mathrm{PF}}
    = 
    \frac{1}{\sqrt{2+2(\cos \theta)^{L}}}
    \qty(
    \ket{\theta}
    -
    \ket{-\theta}
    ),
    \label{eq:|PF_iy>}
\end{align}
where $\ket{\theta}$ is the product state
\begin{align}
    \ket{\theta}
    = 
    \bigotimes_{i=0}^{L-1}
    \qty(
    \cos \frac{\theta}{2}
    + 
    \sin \frac{\theta}{2}
    a_{i}^\dag 
    )
    \ket{0_{i}}.
    \label{eq:|theta_iy>}
\end{align}
Here $\ket{0_{i}}$
is the local fermionic vacuum state at ${i}$-th site on the 1D chain of length $L$. 
Notice that we need to use a cat state as in Eq. \eqref{eq:|PF_iy>} in order to have a Gaussian state.
Indeed $\ket{\mathrm{PF}}$ is the ground state of the 1D version of our Hamiltonian \eqref{eq:H_0} with $\gamma =\frac{\sin^2 \theta}{1+\cos^2 \theta}$ and $ h
    =
    \frac{2|\cos \theta|}{1+\cos^2\theta}$, see Refs. \cite{ktm-82,ms-85}.

The 1D states~\eqref{eq:|theta_iy>} correspond to the tilted ferromagnets that have been employed as the prototypical initial configurations to study the entanglement asymmetry and the quantum Mpemba effect both theoretically~\cite{Ares-2023} and experimentally~\cite{Lata-2024}.

The angle $\theta\in[0,\pi]$ tunes the probability of finding a fermion at the $\mathbf{i}$-th site and, therefore, controls how much the $U(1)$ particle-number symmetry is broken \cite{Ares-2023}.
At $\theta=0$ and $\pi$, the state~\eqref{eq:|theta_iy>} is fully empty or occupied, respectively, and it is symmetric. The symmetry is instead broken when $\theta\in(0,\pi)$. 

From the 1D cat state \eqref{eq:|PF_iy>}, we can build two different types of configurations in the 2D
square lattice that are ground states of the Hamiltonian~\eqref{eq:H_0} for specific subsets of the parameters and for which the R\'enyi entanglement asymmetry exhibits qualitatively different behavior after a quench to the Hamiltonian~\eqref{eq:H}. 
Below we analyze them separately.

\subsubsection{Partially-filled product state I}
\label{subsubsec:PFI}

Let us denote by $\ket{\mathrm{PF}_{i_y}}$ the $\ket{\mathrm{PF}}$ state on the $i_y$ row of the 2D lattice and 
let us consider the configuration
\begin{align}
    \ket{\rm PF_{I}}
    = 
    \bigotimes_{i_y=0}^{L_y-1}
    \ket{\mathrm{PF}_{i_y}}, 
    \label{eq:|PF_I>}
\end{align}
which is a product in the transverse direction of partially filled cat states in the longitudinal one.
Being $\ket{\mathrm{PF}}$ the ground state of a 1D chain, $\ket{\rm PF_{I}}$ is the ground state of a 2D model with vanishing transverse couplings, i.e.
it is the ground state of the Hamiltonian \eqref{eq:H_0} for 
\begin{align}
    \gamma_x = \frac{\sin^2 \theta}{1+\cos^2 \theta},\ 
    h
    =
    \frac{2|\cos \theta|}{1+\cos^2\theta}
    ,\ 
    t_x = 1,\ 
    t_y=\gamma_y = 0. 
    \label{params:PFI}
\end{align}

\begin{figure}
    \raggedright
    \includegraphics[width=0.45\textwidth]{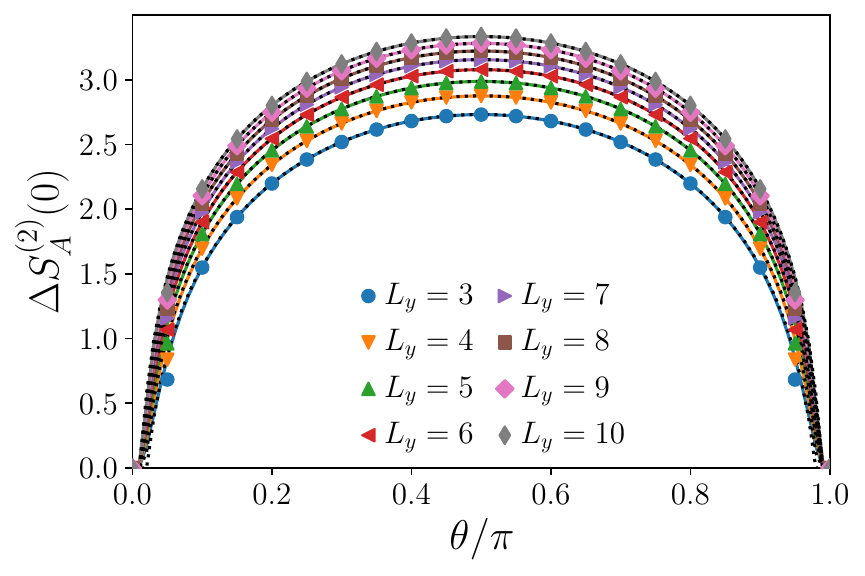}
    \caption{$n=2$ R\'enyi entanglement asymmetry in the partially filled product state I, cf. Eq.~\eqref{eq:|PF_I>}, as a function of the angle $\theta$ for a strip of width $\ell=100$ and several values of the transverse length $L_y$. 
    The solid lines have been obtained using the analytic prediction for the charged moments~\eqref{eq:charged_moment_t=0} in Eq. \eqref{eq:REA_Zn}, the symbols are the exact value of the entanglement asymmetry calculated numerically using Eq.\,\eqref{eq:Z_nny_det}.
    The dashed lines correspond to the saddle point approximation of Eq.~\eqref{eq:REA_PFI_t=0}.  
    }
    \label{fig:PFI_REA_n=2_t=0}
\end{figure}

\par 
The R\'enyi entanglement asymmetry of this state is given by 
Eq.~\eqref{eq:REA_t=0}. Using the 
parameters~\eqref{params:PFI} in Eq.~\eqref{eq:Cooper_pair_density} 
and integrating over $q_x$, we find that the Cooper pair density is 
$\varrho_c=\sin^2\theta/2$. Therefore, 
\begin{align}
    \Delta S_A^{(n)}
    \simeq 
    \frac{1}{2}
    \log
    \frac{ V_A n^\frac{1}{n-1}\pi\sin^2 \theta}{8}. 
    \label{eq:REA_PFI_t=0}
\end{align}
In Fig.~\ref{fig:PFI_REA_n=2_t=0}, we plot Eq.\,\eqref{eq:REA_PFI_t=0} as a function of $\theta$ for the R\'enyi index $n=2$ and several sizes $L_y$ of the transverse direction (solid lines). It agrees well with the exact value (symbols) obtained by evaluating numerically Eq.~\eqref{eq:Z_nny_det}. We observe that the R\'enyi entanglement asymmetry monotonically increases as the angle $\theta$ approaches $\pi/2$, at which the $U(1)$ particle number symmetry is maximally broken, and then monotonically decreases until $\theta=\pi$, angle at which the $U(1)$ particle number symmetry is respected again.

\begin{figure}
    \raggedright
    \includegraphics[width=0.45\textwidth]{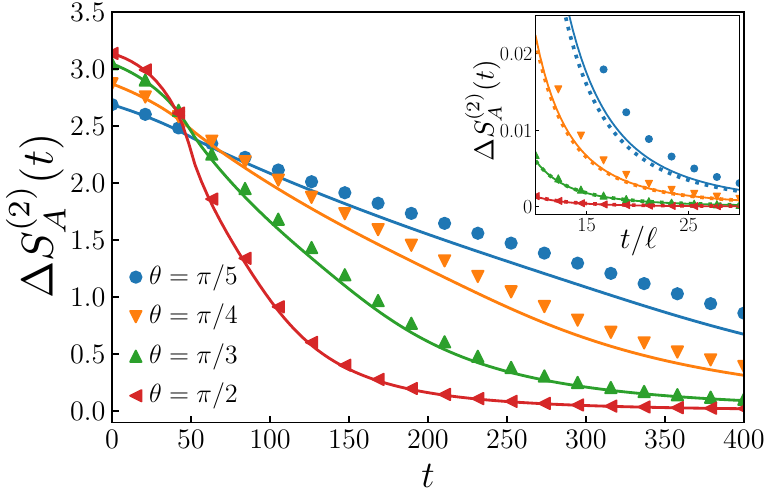}
    \caption{Time evolution of the $n=2$ R\'enyi entanglement asymmetry in the quench from the state \eqref{eq:|PF_I>} for different values of the angle $\theta$. 
    We consider a strip $A$ of size $L_y=8$ and $\ell=100$.
    The solid lines correspond to the ballistic prediction~\eqref{eq:time_ev_REA_Jk} and the symbols are the exact entanglement asymmetry calculated numerically with Eq.~\eqref{eq:Z_nny_det}. 
    The inset is a zoom of the main plot at large times. 
   The dashed lines  correspond to the asymptotic behavior~\eqref{eq:REA_large_t_expand} for large $t/\ell$.  }
    \label{fig:PFI_time}
\end{figure}

We now analyze the time evolution of the R\'enyi entanglement asymmetry after a quench to the Hamiltonian~\eqref{eq:H_0}. 
We can use the analytic expression~\eqref{eq:time_ev_REA_Jk} with the parameters \eqref{params:PFI}. 
We plot the resulting curves in Fig.~\ref{fig:PFI_time} as a function of time
for $n=2$ and several values of the initial angle $\theta$, taking $L_y=8$. We also represent some exact values of the entanglement asymmetry (symbols) computed numerically 
using Eq.~\eqref{eq:Z_nny_det}.
As expected, the R\'enyi entanglement asymmetry tends to zero at large times and the $U(1)$ particle number symmetry broken by the initial state is restored. 
We also observe in Fig.\,\ref{fig:PFI_time} that, for any pair of initial angles $\theta$, the curves described by the R\'enyi entanglement asymmetry cross at a finite time.
This means that, the more the state~\eqref{eq:|PF_I>} breaks the particle number symmetry, the faster the symmetry is restored after the quench.
As we have already said in the introduction, this phenomenon has been 
dubbed quantum Mpemba effect~\cite{Ares-2023}. This plot shows 
that it can also occur in 2D systems and, for the set of 
initial states~\eqref{eq:|PF_I>}, it is robust when we increase the 
length of the transverse dimension $L_y$.
Notice that in Fig.\,\ref{fig:PFI_time} the curves for the smallest tilting angle $\theta$ do not perfectly match with the numerics. This is not surprising since already in 1D it was noted \cite{Murciano-2023} that, when the relaxation is slow, one needs to go to larger values of $\ell$ (and so of $t$) to match the asymptotic results. In the 2D case, this is numerically demanding and does not add any physical insight to our findings. 
This phenomenon will also manifest with certain other initial states and will not be discussed further.

\subsubsection{Partially-filled product state II}
\label{subsubsec:PFII}

From the 1D cat state \eqref{eq:|PF_iy>}, we can also construct the configuration 
\begin{align}
    \ket{\rm PF_{II}}
    = 
    \bigotimes_{i_x=0}^{L_x-1}
    \ket{\mathrm{PF}_{i_x}},
    \label{eq:|PF_II>}
\end{align}
where $\ket{\mathrm{PF}_{i_x}}$ is the $\ket{\mathrm{PF}}$ state on the $i_x$ column of the 2D lattice. 
Hence $\ket{\rm PF_{II}}$ is a product state in the longitudinal direction, 
instead of in the transverse one as $\ket{{\rm PF_I}}$. 
Since $\ket{{\rm PF_I}}$ and $\ket{{\rm PF_{II}}}$ are both built from a non-separable cat state, \eqref{eq:|PF_I>} and \eqref{eq:|PF_II>} are not equivalent. In fact,  \eqref{eq:|PF_II>}
 is the ground state of the Hamiltonian \eqref{eq:H_0} for the set of parameters 
\begin{align}
    t_x=\gamma_x=0,t_y=1,
    \gamma_y = \frac{\sin^2\theta}{1+\cos^2\theta},
    h = \frac{2|\cos\theta|}{1+\cos^2\theta}, 
    \label{params:PFII}
\end{align}
which are different from those in Eq. \eqref{params:PFI}.

\begin{figure}
    \raggedright
    \includegraphics[width=0.45\textwidth]{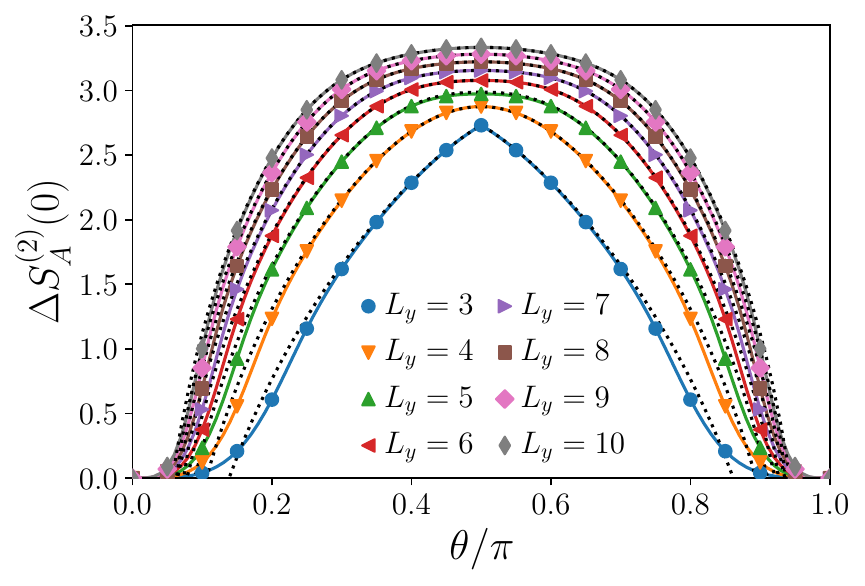}
    \caption{$n=2$ R\'enyi entanglement asymmetry  in the partially-filled product state II, Eq.~\eqref{eq:|PF_II>}, as a function of the angle $\theta$ for a periodic strip $A$ of size $\ell\times L_y$, with $\ell=100$. 
    The solid lines are obtained using the analytic prediction~\eqref{eq:charged_moment_t=0} for the charged moments in Eq. \eqref{eq:REA_Zn}. The symbols are the exact numerical values obtained using \eqref{eq:Z_nny_det}. 
    The dashed curves are the saddle point approximation \eqref{eq:REA_t=0}. 
    }
    \label{fig:PFII_EA_n=2_t=0}
\end{figure}

The R\'enyi entanglement asymmetry of the state $\ket{\mathrm{PF}_\mathrm{II}}$  
behaves as Eq.~\eqref{eq:REA_t=0} for large $\ell$. For $L_y\gg1$, 
we can replace the summation over $n_y$ in 
Eq.~\eqref{eq:Cooper_pair_density} by an integral; in that case, the 
density of Cooper pairs is  $\varrho_\mathrm{c}=\sin^2\theta/2$. 
Therefore, for $L_y$ and $\ell$ large enough, the R\'enyi 
entanglement asymmetries of the states $\ket{\mathrm{PF}_{\mathrm{I}}}$ and $\ket{\mathrm{PF}_{\mathrm{II}}}$ coincide, cf. Eq.~\eqref{eq:REA_PFI_t=0}. For finite $L_y$, the sum over $n_y$ in 
Eq.~\eqref{eq:Cooper_pair_density} cannot be explicitly done.  In 
Fig.~\ref{fig:PFII_EA_n=2_t=0}, we plot the saddle point 
approximation~\eqref{eq:REA_t=0} as a function of the angle $\theta$ 
for several values of $L_y$ (dashed lines) and we compare it with the 
exact values (symbols) obtained using the determinant 
formula~\eqref{eq:Z_nny_det}. We 
have also represented (solid curves) the result of integrating directly in $\boldsymbol{\alpha}$ the analytic prediction of Eq.~\eqref{eq:charged_moment_t=0} for the charge moments without performing a saddle point approximation. In general, we obtain a good agreement between the three approaches. 
As in the other partially-filled product state~\eqref{eq:|PF_I>}, the entanglement asymmetry monotonically increases with $\theta$ until $\theta=\pi/2$, at which it behaves non-analytically when $L_y=3$, and then monotonically decreases.

\begin{figure}
    \raggedright
    \includegraphics[width=0.45\textwidth]{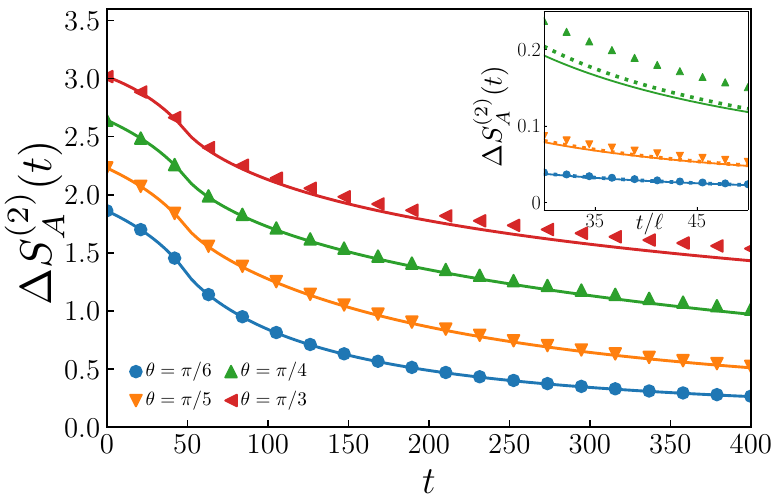}
    \caption{Time evolution of the $n=2$ R\'enyi entanglement asymmetry  after a quench from the state \eqref{eq:|PF_II>}. We choose a system of transverse length $L_y=8$ and as subsystem a periodic strip of width $\ell=100$.
    The solid lines correspond to the analytic prediction \eqref{eq:time_ev_REA_Jk} in the ballistic regime and the symbols are exact entanglement asymmetry obtained by evaluating numerically Eq.\,\eqref{eq:Z_nny_det}.
    In the inset, the dotted curves are the asymptotic expression \eqref{eq:REA_large_t_expand} of the entanglement asymmetry for large times. }
    \label{fig:PFII_time}
\end{figure}

The R\'enyi entanglement asymmetry after the quench to the Hamiltonian~\eqref{eq:H} can be obtained with the analytic prediction~\eqref{eq:time_ev_REA_Jk}. 
In Fig.\,\ref{fig:PFII_time}, we plot it as a function of time (solid lines) for the R\'enyi index $n=2$ and taking several initial angles $\theta$ and $L_y=8$. Apart from the good agreement with the exact numerical values (symbols), we can see that the R\'enyi entanglement asymmetry monotonically tends to zero at large times but the curves do not intersect. Namely, the quantum Mpemba effect does not occur for this set of initial states, unlike for the states $\ket{{\rm PF}_{\rm I}}$ considered in Fig.~\ref{fig:PFI_time}. 

\subsection{Isotropic couplings}
\label{subsec:isotropic}

So far we have studied the time evolution of the R\'enyi entanglement asymmetry for the initial states of Eqs.~\eqref{eq:|PF_I>} and \eqref{eq:|PF_II>}, which represent a very restricted subset of couplings of the Hamiltonian~\eqref{eq:H_0}. Let us consider now as initial configurations the ground state of the Hamiltonian \eqref{eq:H_0} with isotropic hoppings and pairings, 
\begin{align}
    t_x = t_y = 1,\ \gamma_x=\gamma_y=\gamma. 
    \label{params:isotropic}
\end{align}
The R\'enyi entanglement asymmetry for some configurations of this kind is already shown in Fig.\,\ref{fig:REA_isotropic}. 

\begin{figure*}
    \includegraphics[width=0.95\textwidth]{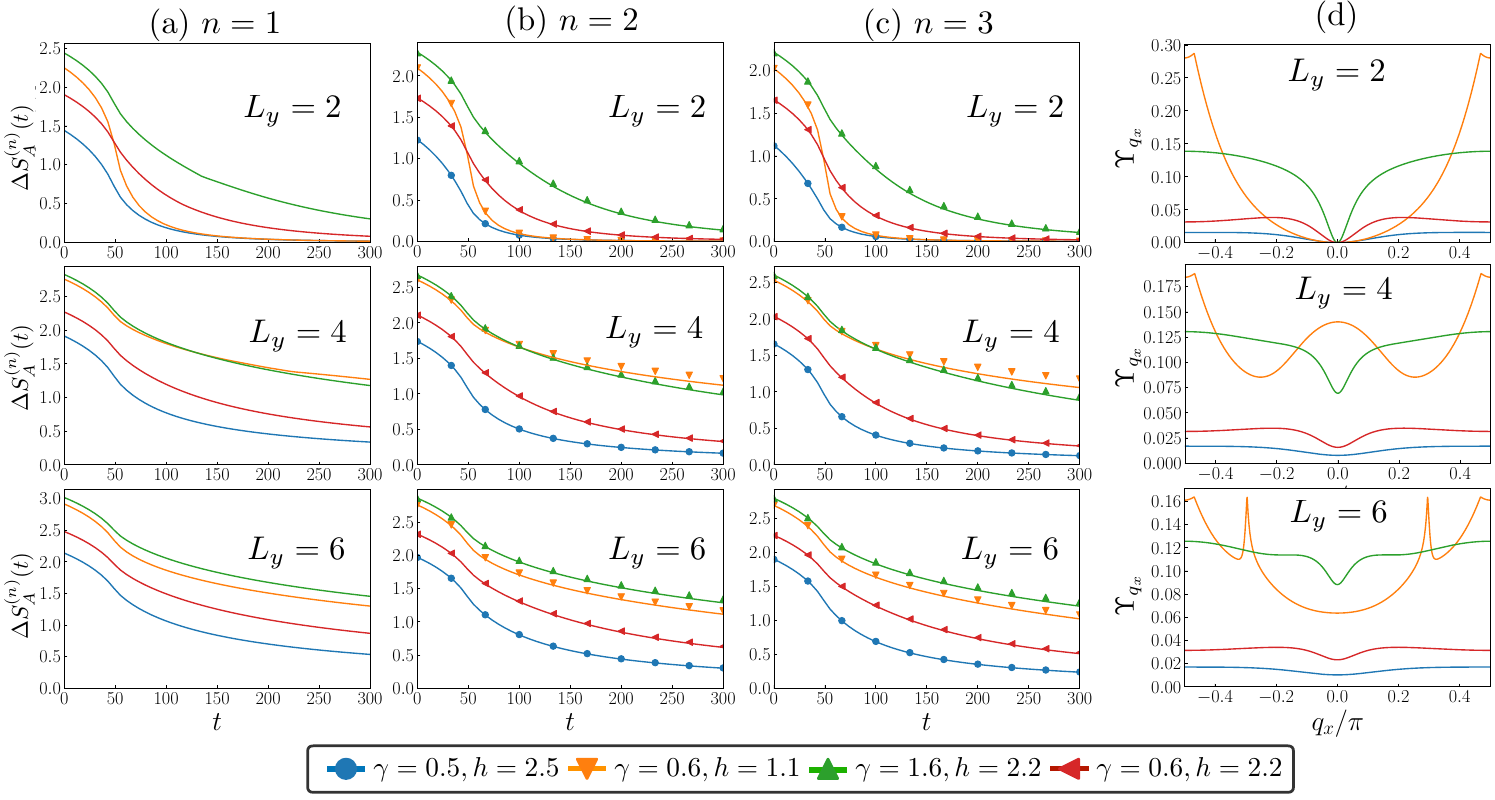}
    \caption{
    (a)-(c) Time evolution of the R\'enyi entanglement asymmetries for R\'enyi indices $n=1$, 2, and 3 after a quench to the Hamiltonian~\eqref{eq:H} starting from different ground states of the Hamiltonian \eqref{eq:H_0} with couplings $t_x=t_y=1$ and $\gamma_x=\gamma_y=\gamma$. 
    The solid lines in (a)-(c) correspond to the analytic predictions \eqref{eq:time_ev_EA_Jk} for $n=1$ and \eqref{eq:time_ev_REA_Jk} for $n=2,3$. 
    The symbols in (b) and (c) are the exact value of the entanglement asymmetry obtained computing numerically the charged moments with the determinant formula~\eqref{eq:Z_nny_det}. We take as subsystem $A$ the periodic strip of Fig.~\ref{fig:system} with $\ell=100$ in all the panels, varying the transverse dimension $L_y$. 
    In panel (d), we represent for the initial states considered in the other panels the function $\Upsilon_{q_x}$ that determines the occurrence of the quantum Mpemba effect according to Eq.~\eqref{ineq:f_final}. 
    }
    \label{fig:ISOTROPIC}
\end{figure*}

In Fig.\,\ref{fig:ISOTROPIC}, we analyze the evolution of the R\'enyi entanglement asymmetry after quenches from the ground state of the Hamiltonian \eqref{eq:H_0} for different couplings satisfying Eq.~\eqref{params:isotropic}.  We consider R\'enyi indices $n=1, 2$ and $3$ (panels (a), (b) and (c) respectively). The solid curves correspond to the analytic prediction of Eq.~\eqref{eq:time_ev_EA_Jk} for $n=1$ and of Eq.~\eqref{eq:time_ev_REA_Jk} for $n=2, 3$,
while the symbols are the exact value calculated numerically using Eq.~\eqref{eq:Z_nny_det}. The qualitative behavior of the entanglement asymmetry does not depend on the R\'enyi index $n$ but it does change as the transverse system length $L_y$ varies. 
For example, for $L_y=2$, the curves corresponding to $(\gamma,h)=(0.6,1.1)$ (orange line) and $(\gamma, h)=(0.6,2.2)$ (red line) 
cross at a certain time and there is quantum Mpemba effect. On 
the other hand, for the same initial states, this phenomenon 
disappears if $L_y=4$ or $6$ since their entanglement asymmetries do 
not cross. It is also remarkable 
the case of the initial states $(\gamma,h)=(0.6,1.1)$ (orange line) and $(\gamma, h)=(1.6,2.2)$ (green line), for which there is no quantum Mpemba effect when $L_y=2$ or $6$, but it occurs in the intermediate size $L_y=4$. 
This observation adds one new player in the Mpemba game: the occurrence of the crossing is not only due to the initial states as in 1D~\cite{Ares-2023, Rylands-2023,Murciano-2023, Lata-2024}, but could depend also on the system geometry.

\subsection{Anisotropic couplings}

\begin{figure*}
    \includegraphics[width=0.95\textwidth]{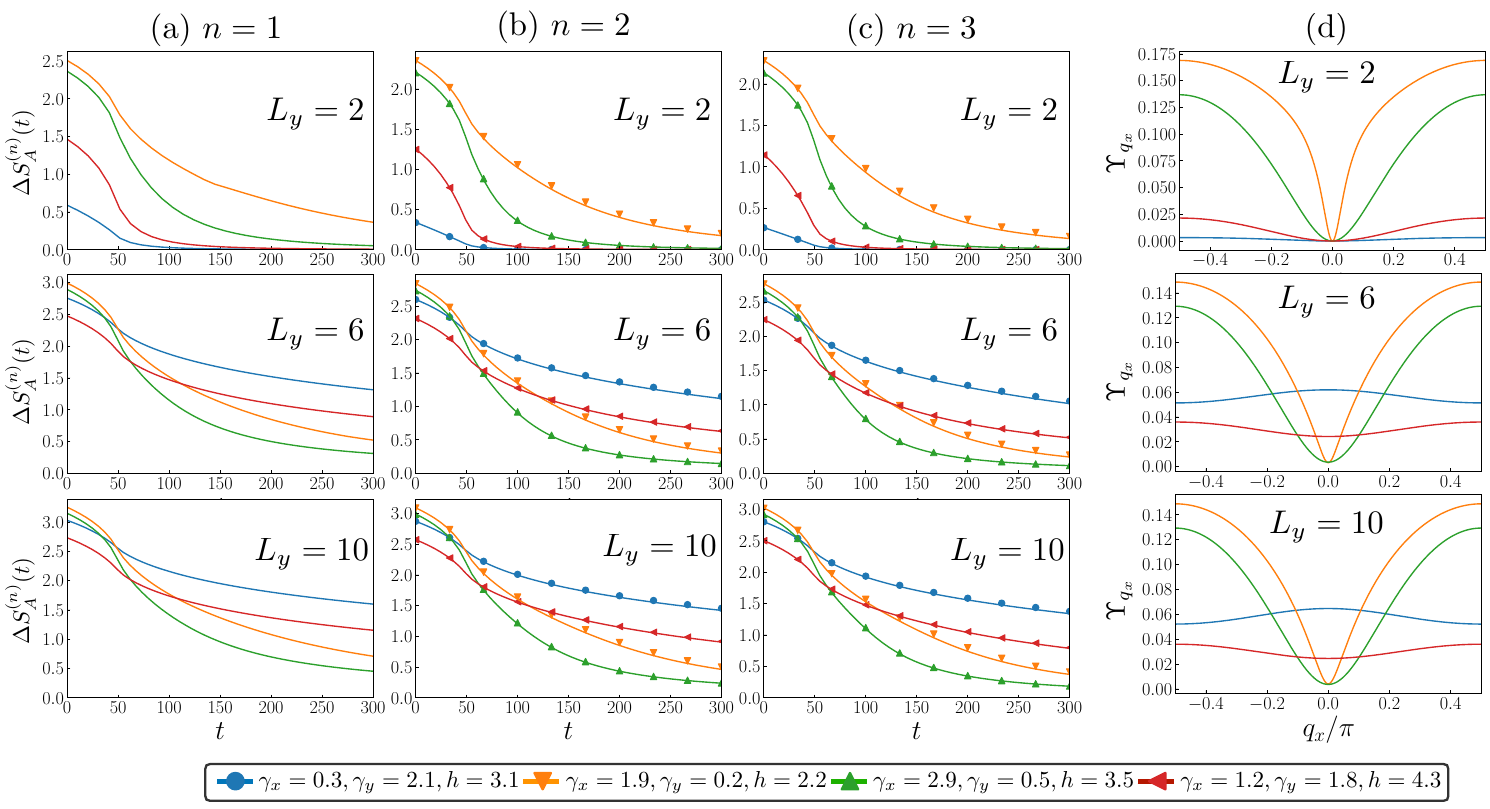}
    \caption{
    (a)-(c) Time evolution of the R\'enyi entanglement asymmetry for $n=1$, $2$, and $3$ after a quench to the Hamiltonian~\eqref{eq:H} starting from the ground state of~\eqref{eq:H_0}, choosing different parameters $t_x=t_y=1$ and $\gamma_x\neq\gamma_y$. 
    The solid lines in panels (a)-(c) correspond to the analytic predictions \eqref{eq:time_ev_EA_Jk} for $n=1$ and \eqref{eq:time_ev_REA_Jk} for $n=2$ and $3$. 
    The symbols in (b) and (c) are the exact value obtained by evaluating numerically Eq.~\eqref{eq:Z_nny_det}. 
     In all the panels, we take a periodic strip of width $\ell=100$ and we vary the transverse system length $L_y$. In panel
    (d), we represent for the initial states studied in the other panels the function $\Upsilon_{q_x}$ that appears in the condition~\eqref{ineq:f_final} for the occurrence of the quantum Mpemba effect.  
    }
    \label{fig:ANISOTROPIC}
\end{figure*}

To show the rich phenomenology of the quantum Mpemba effect in 2D, we 
now consider as initial configurations the ground state of the 
Hamiltonian \eqref{eq:H_0} with anisotropic pairing terms, $\gamma_x\neq\gamma_y$, and isotropic hopping, $t_x=t_y=1$. 
In Fig.\,\ref{fig:ANISOTROPIC} (a)-(c), we plot the time evolution of 
the R\'enyi entanglement asymmetries for several of these states.
The solid curves correspond to the analytic prediction in 
Eq.~\eqref{eq:time_ev_EA_Jk} for $n=1$ (panel (a)) and in 
Eq.~\eqref{eq:time_ev_REA_Jk} for $n=2, 3$ (panels (b) and (c)). The 
symbols are the exact numerical value obtained using the determinant formula~\eqref{eq:Z_nny_det}.
We see a behavior completely opposite to that of Fig.\,\ref{fig:ISOTROPIC}: here there is no quantum Mpemba effect for $L_y=2$, as the curves do not intersect, while, if we increase $L_y$, some of them cross. These crossings are robust for larger values of $L_y$.

\section{Conditions for the quantum Mpemba effect}
\label{sec:condition for QME}
In Sec.\,\ref{sec:time evolution of REA}, we have seen that the quantum Mpemba effect does not always happen for any pair of couplings
of the initial Hamiltonian~\eqref{eq:H_0}. Moreover, whether it takes place or not depends not only on the initial configurations 
but also on the size of the system in the transverse direction, 
$L_y$. In this section, we delve into the reasons for this behavior  
and we derive the microscopic conditions for the quantum Mpemba effect to 
occur from the analytic prediction for the R\'enyi and von Neumann entanglement 
asymmetries in Eqs.~\eqref{eq:time_ev_REA_Jk} and \eqref{eq:time_ev_EA_Jk}. 
\par 
Let us consider the quench from the initial states $\ket{{\rm GS}_{1}}$ and $\ket{{\rm GS}_{2}}$ and assume that the first breaks more the $U(1)$ particle number symmetry than the second. 
In this case, their entanglement asymmetries satisfy 
\begin{align}
    \Delta S_{A, 1}^{(n)}(t=0)
    >
    \Delta S_{A, 2}^{(n)}(t=0).
\end{align}
According to Eq.\,\eqref{eq:REA_t=0}, the inequality above implies that the density of Cooper pairs is larger in the state $\ket{{\rm GS}_1}$ than in $\ket{{\rm GS}_2}$,
\begin{align}
    \varrho_{\mathrm{c},1}
    >
    \varrho_{\mathrm{c},2}. 
    \label{ineq:t=0_Cooper_pair}
\end{align}
This result is intuitive because the condensation of Cooper pairs is induced by the breaking of the $U(1)$ particle number symmetry and thus the number of Cooper pairs increases as more the symmetry is broken. 
\par 
In this setup, the quantum Mpemba effect occurs if and only if there exists a time $t_{\rm M}$ after which the reduced density matrix $\rho_{A,1}(t)$ is always more symmetric than $\rho_{A,2}(t)$, i.e., 
\begin{align}
    \Delta S_{A, 1}^{(n)}(t)
    <
    \Delta S_{A, 2}^{(n)}(t)
    \quad 
    \forall t>t_{\rm M}.
    \label{ineq:t>t*_REA}
\end{align}
Therefore, to determine the conditions under which the inequality \eqref{ineq:t>t*_REA} holds, we need to analyze the behavior of the entanglement asymmetry at large times, i.e. at $\zeta=t/\ell\gg1$. 
\par 
Let us consider the analytic prediction found in Eqs.~\eqref{eq:time_ev_REA_Jk}-\eqref{eq:time_ev_EA_Jk} for the time evolution of the R\'enyi and von Neumann entanglement asymmetries. Since $x_{q_x}(\zeta)\to 0$ when $t\to\infty$, the exponent $A_{1, n_y}(\alpha, \zeta)$  inside the function $J_k(t)$, introduced in Eq.~\eqref{eq:J_k}, tends to zero in that limit. We
can then expand that exponential function and restrict us to the first-order term. Using that approximation, in Appendix~\ref{app:large_times} we find that at large times the R\'enyi entanglement asymmetry behaves as 
\begin{equation}\label{eq:REA_J_0}
\Delta S_A^{(n)}(t)\simeq \frac{n}{1-n}\log J_0(t)
\end{equation}
and, in the case $n\to 1$, as
\begin{equation}\label{eq:vNEA_t>>1}
\Delta S_A^{(1)}(t)\simeq -(1-J_0(t))\log(1-J_0(t)).
\end{equation}
Thus the condition of Eq.~\eqref{ineq:t>t*_REA} for the occurrence of the quantum
Mpemba effect can be recast in terms of $J_0(t)$ as (here $J_{0, 1}$ and $J_{0, 2}$ refer to $J_0$ in the states 1 and 2, respectively)
\begin{equation}\label{eq:J_0_cond}
J_{0, 1}(t)>J_{0, 2}(t),\quad \forall t>t_\mathrm{M},
\end{equation}
which is independent of the R\'enyi index $n$.
This result can be physically interpreted following the reasoning
presented in the 1D case in~\cite{Rylands-2023}. 
The $n=1$ charged moment~$Z_1(\alpha, t)={\rm Tr}(\rho_A(t) e^{\im\alpha Q_A})$ is 
the full counting statistics, the cumulant generating function of the charge $Q_A$~\cite{Bertini-2023-2, Cherng-2007, Groha-2017}. From it, we can extract the probability $P(q)$ of measuring a charge $q$ in the periodic strip $A$ in the initial state from the formula
\begin{equation}\label{eq:prob_p_q}
P(q)=\int_0^{2\pi}\frac{d \alpha}{2\pi} e^{\im\alpha q}Z_1(\alpha, 0).
\end{equation}
For large $\ell$, $Z_1(\alpha, 0)$ behaves as 
\begin{equation}\label{eq:fcs}
Z_1(\alpha, 0)=e^{\ell\sum_{n_y=0}^{L_y-1}A_{1, n_y}(\alpha, 0)}.
\end{equation}
By comparing Eqs.~\eqref{eq:J_k} and \eqref{eq:prob_p_q}-\eqref{eq:fcs}, we can conclude that $J_0(0)$ is precisely the probability of finding a charge $q=0$ in $A$ in the initial state. Observe that, when $t>0$, the function $x_{q_x}(\zeta)$ that appears in the exponent $A_{1, n_{y}}(\alpha, \zeta)$ of $J_k(t)$ vanishes, $x_{q_x}(\zeta)=0$, for the modes with longitudinal velocity $2|v_{q_x}|\zeta>1$; that is, it filters out the contribution of the modes with the fastest group velocity in the longitudinal direction $v_{q_x}$. Therefore, as $t$ increases, only the modes with the slowest longitudinal velocity contribute to $J_0(t)$. This implies that 
\begin{equation}
e^{\ell \sum_{n_y=0}^{L_y-1}A_{1, n_y}(\alpha, \zeta)}\simeq {\rm Tr}(\rho_A(0) e^{\im\alpha Q_{\rm sl}}),
\end{equation}
where $Q_{\rm sl}$ represents the charge of the slowest modes. Thus, for large $t$, $J_0(t)$ can be interpreted as the probability that the longitudinally slowest modes carry zero charge. 
Bringing this picture to the inequality~\eqref{eq:J_0_cond}, we conclude that the quantum Mpemba effect occurs when in the state that breaks more the symmetry there are more modes with the slowest longitudinal velocity transporting zero charge than in the initial state that breaks it less. 

We can determine for which pairs of ground states of the Hamiltonian~\eqref{eq:H_0} the condition~\eqref{eq:J_0_cond} is satisfied, and the quantum Mpemba effect occurs, from their mode occupation of Cooper pairs. When $t\to\infty$, the function $J_0(t)$ behaves as, see Appendix~\ref{app:large_times}, 
\begin{equation}\label{eq:large_time_J_0}
J_0(t)\simeq 1-\ell\int_{0}^{2\pi}\frac{d q_x}{2\pi}x_{q_x}(\zeta)\sum_{n_y=0}^{L_y-1}g_{q_x, q_y},
\end{equation}
where 
\begin{align}
    g_{q_x, q_y}
    = 
    -\frac{1}{2L_y}
    \log\frac{1+\sqrt{1-\sin^2\phi_{\mathbf{q}}}}{2},
    \label{eq:f_qx}
\end{align}
and $\sin^2\phi_{\mathbf{q}}=|\bra{{\rm GS}}a_{\mathbf{q}}a_{-\mathbf{q}}\ket{{\rm GS}}|^2$ is the mode occupation of Cooper pairs in the initial state, as we have already mentioned.

Inserting Eq.~\eqref{eq:large_time_J_0} in~\eqref{eq:REA_J_0} and taking into account that $x_{q_x}(\zeta)\to 0$ when $\zeta\to\infty$, we find that the R\'enyi entanglement asymmetry decomposes at large times as
\begin{equation}\label{eq:REA_t>>1_1}
\Delta S_A^{(n)}(t)\simeq\sum_{n_y=0}^{L_y-1}\Delta S_{A, n_y}^{(n)}(t),
\end{equation}
where $\Delta S_{A, n_y}^{(n)}(t)$ is the contribution from the $n_y$ sector of the transverse modes to the entanglement asymmetry,
\begin{align}
    \Delta S_{A, n_y}^{(n)}(t)
    \simeq
    \frac{n V_A}{n-1}
    \int_0^{2\pi}
    \frac{dq_x}{2\pi}
    x_{q_x}(\zeta) 
    g_{q_x, q_y}.
\end{align}

Since $x_{q_x}(\zeta)=0$ for $2|v_{q_x}|\zeta>1$, the modes that contribute to the entanglement asymmetry at large times are those with the slowest longitudinal group velocity $v_{q_x}$, 
which in our case are those around $q_x=0$ and $\pi$ since $v_x=\sin q_x$.
Therefore, Eq.~\eqref{eq:REA_t>>1_1} can be rewritten as 
\begin{equation}
    \Delta S_{A}^{(n)}(t)
    \simeq 
    \frac{n V_A}{n-1}
    \int_{-q_x^*(\zeta)}^{q_x^*(\zeta)}
    \frac{dq_x}{2\pi}
    (1-2\zeta |v_{q_x}|)\Upsilon_{q_x}
    \label{eq:REA_large_time_final},
\end{equation}
with 
\begin{equation}
\Upsilon_{q_x}=\sum_{n_y=0}^{L_y-1}(g_{q_x, q_y}+g_{q_x+\pi, q_y})
\end{equation}
and~$q_x^*(\zeta)=\arcsin(1/(2\zeta))$. Applying this expression, the condition~\eqref{ineq:t>t*_REA} for $t>t_{\rm M}$ can be re-expressed as 
\begin{align}
    \int_{-q_x^*(\zeta)}^{q_x^*(\zeta)}\!\!
    dq_x \Upsilon_{q_x, 1}
    <
    \int_{-q_x^*(\zeta)}^{q_x^*(\zeta)}\!\!
    dq_x
    \Upsilon_{q_x, 2}.
    \label{ineq:t>t^*_f}
\end{align}

According to its definition in Eq.\,\eqref{eq:f_qx}, the function $g_{q_x, q_y}$ is positive definite and even in $q_x$. 
Therefore, there always exists a large enough time $t_{\rm M}$ at which the integral expression in the inequality~\eqref{ineq:t>t^*_f} can be replaced by 
\begin{align}
    \Upsilon_{q_x, 1}
    <
    \Upsilon_{q_x, 2} \quad \forall 
    q_x
    \in
    \qty[
    -q_x^{*}(t_{\rm M}/\ell),
    q_x^*(t_{\rm M}/\ell)].
    \label{ineq:f_final}
\end{align}
This condition can also be directly derived using~\eqref{eq:large_time_J_0} in the condition~\eqref{eq:J_0_cond} for $J_0(t)$. 
Since $t_M$ can be very large, it is enough that this inequality is satisfied in the neighborhood of $q_x=0$.

Inequalities~\eqref{ineq:t=0_Cooper_pair} and \eqref{ineq:f_final} are the necessary and sufficient conditions to observe the quantum Mpemba effect within our setup. They are a generalization to 2D free fermions
of the ones obtained in Ref.~\cite{Murciano-2023} in the 1D case. 
According to them, the quantum Mpemba effect occurs if in the state that initially breaks less the symmetry and, therefore, contains a smaller number of Cooper pairs [Inequality \eqref{ineq:t=0_Cooper_pair}], the contribution to the asymmetry from the modes with slowest longitudinal velocity (i.e. around $q_x=0$ and $\pi$) is larger [Inequality \eqref{ineq:f_final}] than in the initial state that breaks it more.
\par 
In the panel (d) of Figs.\,\ref{fig:ISOTROPIC} and \ref{fig:ANISOTROPIC}, we plot the function $\Upsilon_{q_x}$ that enters in the inequality \eqref{ineq:f_final} in terms of $q_x$ for the initial states considered in the other panels of those figures. 
We find that, whenever the condition \eqref{ineq:t=0_Cooper_pair} is met for a pair of couplings $(\gamma_x,\gamma_y,h)$ that also satisfies the condition~\eqref{ineq:f_final}, their entanglement asymmetries intersect at a certain time, signaling the occurrence of the quantum Mpemba effect. Since $\Upsilon_{q_x}$ contains the contribution of all the transverse momentum sectors, it depends on
the length $L_y$.
We observe that the form of $\Upsilon_{q_x}$ around $q_x=0$ can change a lot as we vary $L_y$. This explains why the occurrence of the quantum Mpemba effect is strongly affected by modifying the length of the system in the transverse direction.
\par 
In closing this section, it is interesting to obtain the explicit behavior of $\Delta S_A^{(n)}(t)$ at large times. 
This is determined by the behavior $\Upsilon_{q_x}$ close to $q_x=0$, with different power laws if $\Upsilon_0$ vanishes or not. 
Indeed, in Eq.\,\eqref{eq:REA_large_time_final}, by expanding $q^*(\zeta)$ as $q^*(\zeta)\simeq 1/(2\zeta)$ and calculating explicitly the integral,  we find 
\begin{multline}
    \Delta S_A^{(n)}(t)
    \simeq 
    \frac{nV_A}{\pi(n-1)}
    \times 
    \begin{dcases}
        \frac{\Upsilon_{0}''}{192\zeta^3},
        & 
        \stackrel{\displaystyle \gamma_y=0\ {\rm or}\ L_y=2}{ {\rm and}\; h\neq 2},
        \\
        \frac{\Upsilon_0}{4\zeta},
        &
        \mathrm{otherwise},
    \end{dcases}
    \label{eq:REA_large_t_expand}
\end{multline}
where $\Upsilon''_{q_x}=\partial_{q_x}^2 \Upsilon_{q_x}$.

For $n\to1$, combining Eqs.~\eqref{eq:vNEA_t>>1}~and~\eqref{eq:large_time_J_0} and following steps analogous to Eqs.~\eqref{eq:REA_t>>1_1}-\eqref{eq:REA_large_time_final}, we eventually obtain  
\begin{multline}
    \Delta S_A^{(1)}(t)
    \simeq\\
    \frac{V_A}{\pi}
    \times
    \begin{dcases}
        \frac{\Upsilon_{0}''}{192\zeta^3}\log (V_A\zeta^3),
        & \stackrel{\displaystyle \gamma_y=0\ {\rm or}\ L_y=2}{ {\rm and}\; h\neq 2},
        \\
        \frac{\Upsilon_0}{4\zeta}\log(V_A\zeta),
        &\mathrm{otherwise}.
    \end{dcases}
    \label{eq:REA_vN_large_t}
\end{multline}
Eqs.~\eqref{eq:REA_large_t_expand} and \eqref{eq:REA_vN_large_t} show that, for $L_y>2$, the behavior of the R\'enyi entanglement asymmetry at large times changes qualitatively depending on whether $\gamma_y$ is zero or not: 
If $\gamma_y=0$, it decreases as $\zeta^{-3}$ (or $\zeta^{-3}\log \zeta^3$ in the limit $n\to1$), while if $\gamma_y\neq 0$ its decay is algebrically slower, as $\zeta^{-1}$ (or $\zeta^{-1}\log \zeta $ when $n\to1$). 
We check the asymptotic behavior~\eqref{eq:REA_large_t_expand} for the partially-filled product states I ($\gamma_y=0$) and II ($\gamma_y\neq0$) in the insets of Figs.~\ref{fig:PFI_time} and \ref{fig:PFII_time} respectively. 
This difference in the asymptotic behavior means that, if we consider as initial states one with $\gamma_y=0$ and another more symmetric one with $\gamma_y\neq0$ (taking different values of $\gamma_x$ and $h$ for each one), then the symmetry is always restored faster in the first configuration. 
In this case, since their asymmetries follow a power law with different exponents, the phenomenon can be seen as a \textit{strong} quantum Mpemba effect~\cite{Murciano-2023}, in analogy with the classical situation, in which the equilibrium can be reached exponentially faster under certain circumstances~\cite{Klich-2019}. 


\section{Conclusions}
\label{sec:conclusion}
In this paper, we have initiated the study of the entanglement asymmetry and the quantum Mpemba effect in 2D, extending the results obtained in Ref.~\cite{Rylands-2023}, and in particular in Ref.~\cite{Murciano-2023}, for 1D free fermionic chains.
We have considered a periodic 2D lattice of free fermions with nearest-neighbor hoppings and superconducting pairing. This system breaks the $U(1)$ particle number symmetry. We have obtained the corresponding entanglement asymmetry for a periodic strip in the ground state. As in the 1D case~\cite{Murciano-2023, Fossati-2024}, we have found that it grows logarithmically with the size of the strip and with the number of Cooper pairs that the state contains. We have then examined how the entanglement asymmetry evolves after a sudden global quench to a Hamiltonian with no pairing terms, which preserves the number of particles. We have derived analytic expressions for the time-evolved R\'enyi entanglement asymmetry in the ballistic regime and interpreted them by extending the quasi-particle picture developed for 1D systems. 

One important goal of our analysis was to check whether the quantum Mpemba effect occurs after the quench; that is, if the symmetry is restored in the strip faster when the initial state breaks it more. 
We have shown that this phenomenon can occur in our setup but, as in 1D, its occurrence depends on the initial states that we consider. 
In addition, we have found that the transverse size of the system plays a very relevant role. 
The change of this length can obliterate or, on the contrary, give rise to it in a non-trivial way, depending on the initial configurations.

To elucidate the reasons of this behavior, we have analyzed how the entanglement asymmetry reaches the stationary regime. At large times, the symmetry restoration is governed by the quasi-particle pairs that propagate with the slowest velocity in the longitudinal direction. Similarly to (interacting) integrable models in 1D~\cite{Rylands-2023}, we have concluded that the quantum Mpemba effect occurs for a couple of initial states when, in the less symmetric initial configuration, the longitudinally fastest modes carry more charge than in the more symmetric initial state. To determine when this happens, we have obtained the contribution to the entanglement asymmetry of the modes with the slowest longitudinal velocity in terms of the density of Cooper pairs in the initial state, extending the analysis carried out in Ref.~\cite{Murciano-2023} for the 1D chain. Since all the transverse momentum sectors contribute additively, it features an intricate dependence on the system size in the transverse direction, explaining why the quantum Mpemba effect is very sensitive to the second dimension.

A crucial point in our study is the choice of a periodic strip as subsystem. This has allowed us to make use of the dimensional reduction and exploit the 1D results of Refs.~\cite{Rylands-2023, Murciano-2023} to extract exact analytical expressions for the entanglement asymmetry at equilibrium and after the quench. 
This strategy might be useful for calculating the entanglement asymmetry in other $d>1$ systems. 
For example, in $d>1$ spin systems, the spin-wave theory is expected to be valid and, if so, the original Hamiltonian can be approximated as a free-boson model.
For free-boson systems, the techniques to compute the entanglement asymmetry presented in this paper based on dimensional reduction and the Gaussianity of the state are expected to be applicable, as in Refs.~\cite{Song-2011, Frerot-2015}, where the entanglement entropy of the ground state of 2D spin systems is obtained in this way. 
It would also be interesting to choose other subsystems and study the entanglement asymmetry, adapting for example the methods to investigate the entanglement entropy in 2D free fermions of Ref.~\cite{Gibbins-2023}. Of course, it would not be difficult to apply the dimensional reduction employed in this paper, and also in Ref.~\cite{Yamashika-2023} for the entanglement entropy, to other 
closely-related quantities for which there are results in 1D, such as the full counting statistics~\cite{Bertini-2023, Bertini-2023-2, Groha-2017} or the symmetry-resolved entanglement entropies~\cite{Bertini-2023, Parez-2021, Parez-2021-2, Piroli-2022}. 

Our analysis of the quantum Mpemba effect in 2D can also stimulate its investigation in other models, not only theoretically but also experimentally, since the methods used to observe it in a
1D ion trap~\cite{Lata-2024} could be readily implemented in other platforms such as higher dimensional arrays of atoms in optical lattices or superconducting qubits~\cite{Monroe-2021, Daley-2022, Bloch-2008}.

\acknowledgments
We thank C. Rylands for the fruitful discussions. 
SY is supported by Grant-in-Aid for JSPS Fellows (Grant No. JP22J22306). PC and FA acknowledge support from ERC under Consolidator Grant number 771536 (NEMO). 

\appendix

\section{Derivation of Eqs.~\eqref{eq:2d_charged_moments_corr_mat} and \eqref{eq:def of W_n}}\label{app:charged moment}

In this Appendix we present a detailed derivation of Eqs.~\eqref{eq:2d_charged_moments_corr_mat} and~\eqref{eq:def of W_n}. 
The following arguments are based on Ref.~\cite{Murciano-2023}.

To obtain Eq.~\eqref{eq:2d_charged_moments_corr_mat}, we have to evaluate the trace of the product of $\rho_Ae^{\im \alpha_{j,j+1} Q_A}$. Since the reduced density matrix $\rho_A$ is Gaussian, it can be written as 
\begin{align}
    \rho_A 
    = 
    \frac{1}{Z}e^{-\frac{1}{2}\sum_{{\bf i,i'}\in A}{\bf a_i}(H_A)_{\bf i,i'}{\bf a_{i'}^\dag}},
\end{align}
where 
\begin{align}
    Z
    &=
    \Tr (e^{-\frac{1}{2}\sum_{{\bf i,i'}\in A}{\bf a_i}(H_A)_{\bf i,i'}{\bf a_{i'}^\dag}})
\end{align}
is the normalization constant and $H_A$ is the single-particle entanglement Hamiltonian, which in terms of the two-point correlation matrix $\Gamma$ reads~\cite{Peschel-2003}
\begin{align}
    H_A
    =
    -\log\qty(\frac{I+\Gamma}{I-\Gamma}). 
    \label{eq:entanglement Hamiltonian}
\end{align}
The number operator $Q_A$ is quadratic and, therefore, $e^{\im \alpha Q_A}$ can be also cast in the form
\begin{align}
    e^{\im \alpha Q_A}
    = 
    e^{\frac{\im}{2}\alpha \sum_{\bf i \in A} {\bf a_i}\sigma_z {\bf a_i^\dag}}.
\end{align}
Employing the Baker-Campbell-Hausdorff formula, we obtain 
\begin{align}
    \rho_A e^{\im \alpha Q_A}
    = 
    \frac{1}{Z}
    e^{\frac{1}{2}\sum_{\bf i,i'\in A}{\bf a_i}(H_A(\alpha))_{\bf i,i'} {\bf a_{i'}^\dag}},
\end{align}
with $H_A(\alpha)=\log(e^{-{H_A}}e^{\im \alpha I\otimes \sigma_z})$. 
In the same way, one readily finds that 
\begin{align}\label{eq:rho_Ae^Q}
    \prod_{j=1}^{n}
    \rho_A e^{\im \alpha_{j,j+1} Q_A}
    = 
    \frac{1}{Z^n}
    e^{\frac{1}{2}\sum_{\bf i,i'\in A}{\bf a_i}
    (H_A(\boldsymbol{\alpha}))_{\bf i,i'} {\bf a_{i'}^\dag}}, 
\end{align}
with 
\begin{align}
    H_A(\boldsymbol{\alpha})
    = 
    \log(\prod_{j=1}^{n}e^{-H_A}e^{\im \alpha_{j,j+1}I\otimes \sigma_z}).
\end{align}
Finally, applying the trace formula for a generic non-Hermitian matrix $M$~\cite{Fagotti-2010},
\begin{align}\label{eq:Fagotti-formula}
    \Tr(e^{\frac{1}{2}\sum_{\bf i,i'\in A}{\bf a_i}M_{\bf i,i'}{\bf a_{i'}^\dag}})
    = 
    \sqrt{\det(I+e^{M})},
\end{align}
to Eq.~\eqref{eq:rho_Ae^Q} and using Eq.~\eqref{eq:entanglement Hamiltonian}, we obtain 
\begin{equation}
    Z_n(\boldsymbol{\alpha})= 
    Z^{-n/2}
    \sqrt{\det(I+W_n(\boldsymbol{\alpha}))}, 
\end{equation}
where $W_n(\boldsymbol{\alpha})$ was introduced in Eq.~\eqref{eq:def of W_n}.
Finally, rewriting the normalization factor $Z$ in terms of $\Gamma$ using Eqs.~\eqref{eq:entanglement Hamiltonian} and \eqref{eq:Fagotti-formula}, we arrive at Eq.~\eqref{eq:2d_charged_moments_corr_mat}.

\section{Derivation of Eq.~\eqref{eq:REA_t=0}}\label{app:saddle_point}

In this Appendix, we give a detailed derivation of Eq.~\eqref{eq:REA_t=0}. 
Plugging the analytic prediction~\eqref{eq:charged_moment_t=0} for the charged moments into Eq.~\eqref{eq:REA_Zn}, the R\'enyi entanglement asymmetry at $t=0$ is given by the 
$n$-fold integral
\begin{align}
    e^{(1-n)\Delta S_A^{(n)}}
    =
    \!\!\!\!
    \int \limits_{[0,2\pi]^n}
    \!\!\!\!
    \frac{d^n \boldsymbol{\alpha}}{(2\pi)^n}
    e^{\ell \sum_{n_y=0}^{L_y-1}A_{n,n_y}(\boldsymbol{\alpha})}.
    \label{eq:REA_t=0_1}
\end{align}
To evaluate this integral, we introduce the new set of variables 
\begin{align}
    \beta_i = 
    \begin{cases}
        \alpha_1, & {\rm if}\,\, i=0,
        \\
        \alpha_{i+1}-\alpha_i,& \mathrm{otherwise}.
    \end{cases}
\end{align}
In terms of them, Eq.\,\eqref{eq:REA_t=0_1} is written as 
\begin{multline}
    e^{(1-n)\Delta S_A^{(n)}}
    =\\
    \int \limits_{R_{\boldsymbol{\beta}}}  
    \frac{d^n\boldsymbol{\beta}}{(2\pi)^n}
    e^{\ell \sum_{n_y=0}^{L_y-1}
    \qty[
    \sum_{j=1}^{n-1} A_{1,n_y}(-\beta_j) 
    +A_{1,n_y}\qty(\sum_{j=1}^{n-1} \beta_j)]},
    \label{eq:REA_t=0_2}
\end{multline}
where the domain of integration $R_{\boldsymbol{\beta}}$ is given by the conditions 
\begin{align}\label{eq:R_beta}
    R_{\boldsymbol{\beta}}\ :\ 
    0\leq\sum_{j=0}^i \beta_j \leq 2\pi, \quad i=0,\dots,n-1.
\end{align}
Since the integrand in Eq.~\eqref{eq:REA_t=0_2} is independent of $\beta_0$, the integral over this variable can be easily done.
This results in 
\begin{multline}
    e^{(1-n)\Delta S_A^{(n)}}=\\
    \!\!\!\!\!\!
    \int \limits_{[-2\pi,2\pi]^n}  
    \!\!\!\!\!\!
    \frac{d^{n-1}\boldsymbol{\beta}}{(2\pi)^n}
    \mu(\boldsymbol{\beta})
    e^{\ell \sum_{n_y=0}^{L_y-1}\qty[
    \sum_{j=1}^{n-1} A_{1,n_y}(-\beta_j) 
    +A_{1,n_y}\qty(\sum_{j=1}^{n-1} \beta_j)]},
    \label{eq:REA_t=0_3}
\end{multline}
where
\begin{multline}
    \mu(\boldsymbol{\beta}) 
    = 
    \max\bigg[
    0,2\pi-\max_{i\in[0,n-1]}\qty(\sum_{j=1}^{i}\beta_j)
    \nonumber\\
    +
    \min_{i\in[0,n-1]}\qty(\sum_{j=1}^{i}\beta_j)
    \bigg]
\end{multline}
is the measure of the domain of $\beta_0$ for the constraints in Eq.~\eqref{eq:R_beta}. 
\par 
Since the exponent of the integrand in Eq.\,\eqref{eq:REA_t=0_3} is proportional to $\ell$, the integral over $\boldsymbol{\beta}$ in Eq.\,\eqref{eq:REA_t=0_3} can be evaluated when $\ell\gg 1$ with the saddle point approximation \cite{Daniels-1954}. 
The saddle points of the integrand can be obtained by solving the saddle point condition
\begin{align}
    \frac{\partial}{\partial\beta_j}
    \sum_{n_y=0}^{L_y-1}\qty[\sum_{j=1}^{n-1} A_{1,n_y}(-\beta_{j})+A_{1,n_y}\qty(\sum_{j=1}^{n-1}\beta_j)]
    =0,
\end{align}
with $j=1,...,n-1$.
By solving this set of equations, one finds that there exist $3^{n-1}$ saddle points $\boldsymbol{\beta}^*=(\beta_1^*,\dots,\beta_{n-1}^*)$, with $\beta_j^*=0,\pm \pi$. 
Among them, there are $2^n-2$ saddle points at which $\mu(\boldsymbol{\beta}^*)=\pi$, e.g. $\beta_j^*=\pi\delta_{j,1}$, and one saddle point $\boldsymbol{\beta}^*=(0, \dots, 0)$ at which $\mu(\boldsymbol{\beta}^*)=2\pi$. 
Around these saddle points, the exponent of the integrand in Eq.\,\eqref{eq:REA_t=0_3} has the same expansion at quadratic order of $\boldsymbol{\beta}^*$. For the rest of the saddle points, $\mu(\boldsymbol{\beta}^*)=0$. 
Taking into account all these factors, the right-hand side of Eq.\,\eqref{eq:REA_t=0_3} can be approximated as 
\begin{equation}
    e^{(1-n)\Delta S_A^{(n)}}
    \simeq
    2^{n-1}
    \!\!\!\!\!\!\!\!
    \int \limits_{[-\infty,\infty]^{n-1}}
    \!\!\!\!\!\!\!\!
    \frac{d^{n-1}\boldsymbol{\beta}}{(2\pi)^{n-1}}
    e^{-\frac{V_A\varrho_\mathrm{c}}{2}\sum_{i\leq j}\beta_i\beta_j}, 
    \label{eq:REA_t=0_4}
\end{equation}
where $\varrho_{\rm c}$ is given in Eq.\,\eqref{eq:Cooper_pair_density} and $V_A=\ell L_y$. This is a Gaussian integral that can be calculated with the standard formulas. We then arrive at Eq.\,\eqref{eq:REA_t=0}.

\section{Derivation of Eqs.~\eqref{eq:time_ev_REA_Jk} and \eqref{eq:time_ev_EA_Jk}}
\label{app:vN limit}

In this Appendix, we describe how to obtain Eqs.~\eqref{eq:time_ev_REA_Jk} and \eqref{eq:time_ev_EA_Jk} of the main text following the steps  presented in Ref.~\cite{Rylands-2023} for the 1D case. 

We start by rewriting the moments of the symmetrized reduced density matrix $\rho_{A,Q}(t)$ using the variables $\beta_j=\alpha_{j+1}-\alpha_j$, $j=1,\dots, n$,
\begin{gather}
    \Tr[\rho_{A,Q}(t)^n]
    = 
    \!\!\!
    \int \limits_{[0,2\pi]^n}
    \!\!\!
    \frac{d^n \boldsymbol{\beta}}{(2\pi)^{n-1}}
    \delta \qty(\sum_{j=1}^n \beta_j)
    Z_n(\boldsymbol{\beta},t), 
    \label{eq:moments_rho_AQ_dirac}
\end{gather}
where $\delta(x)$ is the $2\pi$-periodic Dirac delta function.

The time evolution of the charged moments is given by Eq.~\eqref{eq:2D_time_ev_charged_moments}. 
Changing variables to  $\beta_j$ and plugging the obtained result in Eq.~\eqref{eq:moments_rho_AQ_dirac}, we have 
\begin{multline}
    e^{(1-n)\Delta S_A^{(n)}}=  \\
    \int \limits_{[0,2\pi]^n} \!\!\!\!
    \frac{d^n \boldsymbol{\beta}}{(2\pi)^{n-1}}
    \delta \qty(\sum_{j=1}^n\beta_j)
    e^{
    \ell \sum_{j=1}^n \sum_{n_y=0}^{L_y-1}A_{1,n_y}(\beta_j,\zeta)
    }.
    \label{eq:REA_vN_1}
\end{multline}
Using the Poisson summation formula for the $2\pi$-periodic Delta function,
\begin{align}
    \delta \qty(\sum_{j=1}^n \beta_j)
    =
    \frac{1}{2\pi}
    \sum_{k=-\infty}^\infty
    \exp(\im k \sum_{j=1}^n \beta_j), 
\end{align}
we directly obtain Eq.~\eqref{eq:time_ev_REA_Jk}.
To find the von Neumann entanglement asymmetry~\eqref{eq:time_ev_EA_Jk}, we have to take the analytic continuation $n\mapsto z\in\mathbb{R}$ in Eq.~\eqref{eq:time_ev_REA_Jk}. Since in our case the functions $J_k(t)$ are real~\cite{Rylands-2023}, the analytic continuation $J_k^n\mapsto e^{z\log J_k}$ is unique according to the Carlson theorem~\cite{Boas-1954}. Taking it in~\eqref{eq:time_ev_REA_Jk} and calculating the limit $z\to 1$ using L'Hôpital rule, we directly find~\eqref{eq:time_ev_EA_Jk}.

\section{Entanglement asymmetry at large times}\label{app:large_times}
In this Appendix, we derive the asymptotic behavior at large times of the R\'enyi and von Neumann entanglement asymmetries reported in Eqs.~\eqref{eq:REA_J_0} and \eqref{eq:vNEA_t>>1} respectively.

In Eqs.~\eqref{eq:time_ev_REA_Jk} and \eqref{eq:time_ev_EA_Jk}, we have obtained the time evolution of the R\'enyi and von Neumann entanglement asymmetries in terms of the function $J_k(t)$ defined in Eq.~\eqref{eq:J_k}. Since $x_{q_x}(\zeta)\to 0$ when $t\to\infty$, the exponent $A_{1, n_y}(\alpha, \zeta)$ in the integrand of $J_k(t)$ becomes small at large times. Therefore, in that limit, we can 
expand the exponent in $J_k(t)$ and restrict us to the first order term,
\begin{equation}\label{eq:J_k_large_t_exp_0}
J_k(t)\simeq\delta_{k, 0}+\ell\int_{0}^{2\pi}\frac{d\alpha}{2\pi}e^{\im \alpha k}\sum_{n_y=0}^{L_y-1}A_{1, n_y}(\alpha,\zeta).
\end{equation}
Taking in this expression $k=0$ and performing the 
integral in $\alpha$ we obtain Eq.~\eqref{eq:large_time_J_0} in the main text.

If we replace $A_{1, n_y}(\alpha, \zeta)$ in Eq.~\eqref{eq:J_k_large_t_exp_0} by its explicit expression~\eqref{eq:A_n_ny_t} and we take into account that $x_{q_x}(\zeta)$ is zero for the modes that satisfy $2\zeta |v_{q_x}|>1$ with $v_{q_x}=\sin q_{x}$, then
\begin{multline}
J_k(t)\simeq \delta_{q, 0}+\ell\int_{0}^{2\pi}\frac{d\alpha}{2\pi}e^{\im\alpha k}
\int_{-\frac{1}{2\zeta}}^{\frac{1}{2\zeta}}\frac{dq_x}{4\pi}(1-2\zeta|v_{q_x}|)\\
\times\sum_{n_y=0}^{L_y-1}(f_{q_x, q_y}(\alpha)+f_{q_x+\pi, q_y}(\alpha)).
\end{multline}
Expanding the integrand around $q_x=0$ up to second order and calculating explicitly the integral in $q_x$, we find that
\begin{align}
    J_k (t)
    \simeq
    \delta_{k,0}
    + 
    d_k X(\zeta),  
    \label{eq:J_k_dX}
\end{align}
where 
\begin{align}
    X(\zeta) 
    &= 
    \begin{dcases}
        \frac{1}{384\pi \zeta^3},
        & 
        \stackrel{\displaystyle \gamma_y=0\ {\rm or}\ L_y=2}{ {\rm and}\; h\neq 2},
        \\
        \frac{1}{8\pi \zeta},
        &
        {\rm otherwise},
    \end{dcases}
\end{align}
\begin{align}
    d_k 
    &= 
    \ell 
    \int \limits_0^{2\pi}
    \frac{d\alpha}{2\pi}
    e^{\im k \alpha}
    \nonumber\\
    &\times 
    \begin{dcases}
        \sum_{n_y=0}^{L_y-1}(f_{0, q_y}''(\alpha)+f_{\pi, q_y}''(\alpha)),
        & 
        \stackrel{\displaystyle \gamma_y=0\ {\rm or}\ L_y=2}{ {\rm and}\; h\neq 2},
        \\
        \sum_{n_y=0}^{L_y-1}\left(f_{0, q_y}(\alpha)+f_{\pi, q_y} (\alpha)\right),
        &
        {\rm otherwise},
    \end{dcases}
\end{align}
with $f_{\bf q}''(\alpha)=\partial_{q_x}^2 f_{\bf q}(\alpha)$.

At large time $\zeta\to\infty$,  $X(\zeta)\to 0$ and so the contribution to the R\'enyi entanglement  asymmetry~\eqref{eq:time_ev_REA_Jk} of the terms $k\neq 0$ is sub-leading with respect to the term
$k=0$. 
Consequently, plugging~\eqref{eq:J_k_dX} into~\eqref{eq:time_ev_REA_Jk} and expanding in $X(\zeta)\ll 1$, we find  the result in the main text
\begin{equation}
\Delta S_A^{(n)}\simeq 
\frac{n}{1-n}\log J_0(t)
\simeq 
\frac{n}{1-n}d_0 X(\zeta).
\end{equation}

In the case of the von Neumann entanglement asymmetry, plugging Eq.\,\eqref{eq:J_k_dX} into Eq.\,\eqref{eq:time_ev_EA_Jk} and expanding in terms of $X(\zeta)\ll 1$, we obtain 
\begin{align}
    \Delta S_A^{(1)}(t)
    &\simeq 
    -(1-J_0(t))\log(1-J_0(t))\\
    &\simeq 
    d_0X(\zeta) \log(-d_0 X(\zeta)),
\label{eq:REA_vN_expand}
\end{align}
where we neglected the sub-leading terms in $X(\zeta)$.

\end{document}